\documentclass[usenatbib]{mnras}
\usepackage[T1]{fontenc}
\usepackage{ae,aecompl}
\usepackage{amsmath}
\usepackage{amssymb}
\usepackage{mathtools}
\usepackage{bm}
\usepackage{cleveref}

\usepackage{graphicx}
\usepackage{caption}
\usepackage{subcaption}
\usepackage{txfonts}
\usepackage{hyperref}
\usepackage{multirow}
\usepackage{hhline}
\title[Extra Galactic Globular Clusters. Preliminary results]{MOCCA Survey Database: Extra Galactic Globular Clusters. I.  Method and first results}

\author[A. Leveque, M. Giersz \& M. Paolillo]{A.~Leveque$^{1}$\thanks{E-mail:agostino@camk.edu.pl}, M.~Giersz$^{1}$ \& M. Paolillo$^{2,3}$\\
  $^{1}$ Nicolaus Copernicus Astronomical Center, Polish Academy of Sciences, ul. Bartycka 18, PL-00-716 Warsaw, Poland\\
  $^{2}$ Dip. di Fisica “E. Pancini”, Università di Napoli Federico II, C.U. di Monte Sant’Angelo, Via Cintia, 80126 Naples, Italy\\
  $^{3}$ INFN, Sez. di Napoli, via Cintia, 80126, Napoli, Italy
             }
           
\graphicspath{{Figures/}}

\begin{document} 



   \date{}
   \pagerange{\pageref{firstpage}--\pageref{lastpage}} \pubyear{2019}
   \maketitle

 
  \begin{abstract}
  Over the last few decades, exhaustive surveys of extra Galactic globular clusters (EGGCs) have become feasible. Only recently, kinematical information of  globular clusters (GCs) were available through Gaia DR2 spectroscopy and also proper motions. 
  On the other hand, simulations of GCs can provide detailed information about the dynamical evolution of the system. We present a preliminary study of EGGCs' properties for different dynamical evolutionary stages. We apply this study to 12 Gyr-old GCs simulated as part of the MOCCA Survey Database. Mimicking observational limits, we consider only a subssample of the models in the database, showing that it is possible to represent observed Galactic GCs. In order to distinguish between different dynamical states of EGGCs, at least three structural parameters are necessary. The best distinction is achieved by considering the central parameters, those being observational core radius, central surface brightness, ratio between central and half-mass velocity dispersion, or similarly considering the central color, the central V magnitude and the ratio between central and half-mass radius velocity dispersion, although such properties could be prohibitive with current technologies. A similar but less solid result is obtained considering the average properties at the half-light radius, perhaps accessible presently in the Local Group. Additionally, we mention that the color spread in EGGCs due to internal dynamical models, at fixed metallcity, could be just as important due to the spread in metallicity. 
  
  
   
\end{abstract}

\begin{keywords}
globular clusters: general
\end{keywords}

%

\section{Introduction}

Star clusters are important natural laboratories for probing galaxy formation and evolution, stellar dynamics, and for testing stellar evolution theory, such as the physical nature of ``exotic'' objects. 

From observations, it is clear that globular clusters (GCs) are very common for all types of galaxies. They can provide a powerful diagnostic for galaxy formation, star formation in galaxies, galaxy interaction and mergers, and the distribution of dark matter in galaxies. Indeed, galaxy-galaxy interactions can trigger major star-forming events and the formation of massive star clusters. The properties of GCs systems in various galaxies can constrain the formation and the evolution of their host galaxies. In particular, Galactic GCs can be used to constrain our Galaxy's halo structure and its formation history.

Our knowledge about GCs increased dramatically over the last couple decades thanks to an enormous amount of very detailed observational data. By taking advantage of the \emph{Hubble Space Telescope} and the newest large ground-based telescopes, a detailed study of extra Galactic globular clusters (EGGCs) has become possible (\citet{Larsen2001}; \citet{Cote2004}; \citet{Peng2006,Peng2008}; see \citet{Brodie2006,Kruijssen2014} and \citet{Renaud2020}  and references therein). The main discovery in recent decades is the bimodality of GCs in the color distribution, a common feature in all type of galaxies
(\citet{Zepf1993}; \citet{Ostrov1993}; \citet{Kundu2001}; but see also \citet{Cantiello2007}). 
The $V-I$ color distribution for bright early-type galaxies usually show a blue peak at $V-I = 0.95 \pm 0.02$, corresponding to $[Fe/H] \sim -1.5$ and a red peak at $V-I = 1.18 \pm 0.04$, corresponding to $[Fe/H] \sim -0.5$ \citep{Larsen2001}. 

\cite{Peng2006} found, as part of the ACS Virgo Cluster Survey \citep{Cote2004}, a relation between host galaxy luminosity (and mass) and GC metallicity for metal-poor GCs in their study of early-type galaxies, which may suggest a universal enrichment during the formation of both the metal-rich and metal-poor populations. Moreover, a larger color dispersion for metal-rich GCs was found, that is metal-rich GCs have nearly twice the color dispersion as metal-poor GCs. Finally, metal-rich GCs show an average half-light radius  smaller than the metal-poor GCs \citep{Jordan2005}. An increase of the half-light radius with galactocentric distance has been found (at least in the central regions of the galaxies, see \cite{Puzia2014}), but with a shallower relationship when compared to Galactic GCs (the slope is $\sim 0.07$ compared to $\sim 0.3$ for Galactic GCs, see \cite{Jordan2005}).

Different scenarios have been suggested to explain the observed color distribution (\citet{Ashman1992}; \citet{Forbes1997}; \citet{Cote1998}), although there is no consensus concerning the its origin. At least two star-formation events in the histories of such galaxies has to be invoked to generate such a bimodality, which can be triggered by major mergers \citep{Ashman1992} or occur in isolation \citep{Forbes1997}. Another explanation could be, in the hierarchical scenario, the accretion of metal-poor GCs from lower-mass galaxies to more massive galaxies, with the metal-rich GCs created in situ \citep{Forbes1997,Harris1999}. This scenario has received support recently from the discovery of large populations of intra-cluster GCs  (see for instance \citep{DAbrusco2016,Cantiello2018,Cantiello2020}. 

The knowledge of the dynamical state of EGGCs is even less established. The reason is that the spatial resolution of even the best telescopes is not enough to resolve the internal structure of distant GCs. Some of those GCs could be dissolving, in others an intermediate-mass black hole (IMBH) could be present, and others could be undergoing core collapse
. Different GC dynamical states can reflect their different observational properties. Even if the observation cannot give such information, simulations can provide all the internal and structural properties of the system.
Even if such studies may be available to Galactic GCs, they could not study all types of environments and evolutionary stages of the clusters that we see in outer galaxies, since the Milky Way is a Late Type galaxy, in the relatively small Local Group.

Simulations show that GC systems containing a large number of black holes (BHs) are characterized by a large stellar core and half-light radii and low surface brightness values (\citet{Askar2018}; \citet{Askar&Askar2019}). On the other hand, it is expected that systems containing an IMBH should be characterized by a high central velocity dispersion and high central surface brightness values. However, it is not easy to determine, using global observational properties, whether a particular GC could contain an IMBH, a BH subsystem (BHS), or neither (\citet{Askar2018}; \citet{ArcaSedda2018}, and reference therein).

In this series of papers our goal is to find a correlation between the global properties of EGGCs and their internal dynamical state and to find some observational properties that would help us to distinguish between these dynamical states. Characteristics that differentiate between GC-galaxy interactions and internal GCs' dynamics could be crucial for the study of EGGCs and the history of their host galaxy. For this reason, it would be important to investigate the correlations between the internal dynamics and the observed global properties of GCs. Is the observed bimodality color distribution due only to the spread in metallicity, or can the dynamical evolution also play a role? If yes, how important could it be? How important is the internal dynamical evolution for other observed parameters? Could the internal dynamical evolution influence the correlations between distance from host galaxies (such as size; \citep{Jordan2005,Puzia2014}) and the GC's properties?

In  order to verify if our project goals are achievable, we will firstly compare our model dataset with Milky Way GCs (MWGCs), since it is easier to observer them and to determine their more detailed properties. Successively, we will try to apply our method to EGGCs, setting a distance limit for which our approach would be valid. This first paper is a proof of concept: we will identify a sub-sample in our dataset that would mimic the observational limit of EGGCs, and immediately try to identify some global features in our dataset. In the following papers, we would like to populate an external galaxy with its GCs population (with properties according to observed distributions) using models from the MOCCA-Survey Database and to apply the procedure described in this paper to our simulated EGGC population.


The structure of the paper is as follows: in \hyperref[sec:Models]{section 2}, we provide information on GC simulation models that were used in this study and we describe the method used to determine global parameters of the GCs and the selection of the models by comparison with the observations of Galactic GCs (\cite{Harris1996}, updated 2010). In \hyperref[sec:Results]{section 3} we present the results and in \hyperref[sec:Discussion]{section 4} we discuss the results and in \hyperref[sec:Summary]{section 5} we provide our final conclusions and describe our future work.

\section{Models} \label{sec:Models}
For the scope of this work, we use the results from the MOCCA-Survey Database  \citep{Askar2017} carried out using the MOCCA code \citep{Hypki2013}. The code simulates and follows the long-term dynamical evolution of spherically symmetric stellar clusters, based on H\'enon's Monte Carlo method (\citet{Henon1971}; \citet{Stodolkiewicz1982,Stodolkiewicz1986}; \citet{Giersz2013} and references therein for details about MOCCA code). Stellar and binary evolution are followed using the prescription from the SSE/BSE codes (\citet{Hurley2000,Hurley2002}), whereas the strong interactions (binary-binary and binary-single) are handled by the FEWBODY code \citep{Fregeau2004}. Escaping stars from tidally limited clusters are treated as described in \cite{Fukushige2000}.

The MOCCA-Survey Database \citep{Askar2017} consists of nearly 2000 real star cluster models that span a wide range of initial conditions, provided in Table 1 in \cite{Askar2017}. 
For half of the simulated models, supernovae (SNe) natal kick velocities for NSs and BHs are assigned according to a Maxwellian distribution, with velocity dispersion of $265 \,\,km \,s^{-1}$ \citep{Hobbs2005}. In the remaining cases, BH natal kicks were modified according to the mass fallback procedure described by \cite{Belczynski2002}. Metallicities of the models were selected as follows: $Z = 0.0002,\, 0.001,\, 0.005,\, 0.006, \,0.02$. All models were characterized by a \cite{Kroupa2001} initial mass function (IMF), with a minimum and maximum initial stellar mass of $0.08$ and $100 \,M_\odot$, respectively. The GC models were described by \cite{King1966} with central concentration parameter values $W_0 = 3, 6, 9$. They had tidal radii ($R_t$) equal to: $30, 60, 120 pc$, while the ratios between $R_t$ and half-mass radius ($R_h$) were $50, 25$ or the model was tidally filling 
. The primordial binary fractions were chosen to be $5\%,\, 10\%,\, 30\%,\, 95\%$. Models characterized by an initial binary fraction equal to or lower than 30 per cent had their initial binary eccentricities selected according to a thermal distribution \cite{Jeans1919}, the logarithm of the semi-major axes according to a flat distribution, and the mass ratio according to a flat distribution. For models containing a larger binary fraction, the initial binary properties were instead selected according to the distribution described by \cite{Kroupa1995}, so-called eigen-evolution and mass feeding algorithms. The models consist of $4 \cdot 10^4,\, 1 \cdot 10^5,\, 4 \cdot 10^5, \,7 \cdot 10^5,\,1.2 \cdot 10^6$ objects (stars and binaries). The GCs were assumed to move on a circular orbit at Galactocentric distances between 1 and $50\,\, kpc$. The Galactic potential was modelled as a simple point-mass, taking as central mass the value of the Galaxy mass enclosed within the GC's orbital radius. The GC rotation velocity was set to $220\,\, km\, s^{-1}$ for the whole range of galactocentric distances. As was shown in \cite{Askar2017}, MOCCA models reproduce some observational properties Milky Way GCs quite well. However, even though the simulated GCs experiences a Galactic-like tidal field, they can  be considered to belong to other galaxies, once the correction for galaxy's mass, distance from the galaxy and the rotation velocity are applied, in order to keep the strength of the tidal field like in the MW.

The simulations present in the MOCCA-Survey Database \citep{Askar2017} project include, as output, snapshots of data containing the details for all stars and binaries in the cluster model. These are produced periodically during the cluster evolution and consist of, at a given time, information of spatial, kinematic and stellar evolution properties of each stars (such as positions from the center of the clusters, velocities, mass, radius), plus binary parameters (such as semimajor axis and eccentricity) for each binary in the system. Information about the global properties of the GCs and about each star and binary in the system is available for such simulations. 

\subsection{Projection and photometry}
As a first step, we projected the positions and velocities of stars from each snapshot on the plane of the sky. For binaries, we additionally computed the barycentric orbit of each star using the semimajor axis and eccentricity of the orbit and the mass of each star in the binary center of mass. In this way, we can treat the binary as resolved, considering positions and velocities of single stars in the binary, or unresolved, considering position and velocity of the binary center of mass: this will be useful to determine the importance of resolving binary systems to the integral properties of observed Galactic GCs and the determination of the velocity dispersion profile (VDP). In this work, we will consider only MOCCA models at the $12 \,Gyr$ snapshot and unresolved binaries\footnote{For the comparison with MWGCs we used only 12 Gyr snapshots. In the next papers in the series we will also use 6 and 9 Gyr snapshots for EGGCs.}.

The absolute magnitude for each star has been calculated and assigned using the FSPS code (\citet{Conroy2009}; \citet{Conroy2010}). FSPS is a stellar population synthesis code, returning the integrated spectra and the luminosity in different bands (for example, Johnson-Cousins, HST WFPC2, HST ACS, etc.) for a given stellar population. The main advantage of this software is its flexibility, because the user can choose their preferred set of isochrones and stellar spectral libraries (see \citet{Conroy2010} for more details). Even though the FSPS code is principally aimed to study galaxies and their stellar and dust content through  their observed photometry and  spectral energy distributions, we modify the code in order to obtain  the integrated absolute magnitude of the entire GC, summing up all stars' magnitudes in the system. We obtained also the integrated absolute magnitude of the system at different cluster radii (radii containing $1\%$, $10\%$, $50\%$, $70\%$ and $100\%$ of the total luminosity and at the observational core radius\footnote{In this paper we defined the observational core radius as the distance from the center where the average surface brightness is half of the central surface brightness, and the half-light radius ($R_{hl}$) as the distance from the center of the cluster contains half of the full cluster luminosity.}  $R_c$). In our study, we considered only the filters commonly used to observe GCs, that are: from Johnson-Cousins system U,B,V,R,I; from SDSS: u,g,r,i,z; from HST WFCP2: F255W, F300W, F336W, F439W, F450W, F555W, F606W, F814W, F850LP; from HST ACS F435W, F474W, F555W, F606W, F625W, F775W, F814W, F850LP; from HST WFC3-UVIS F218W, F225W, F275W, F336W, F390W, F438W, F475W, F555W, F606W, F775W, F814W, F850LP; from HST WFC3-IR F098M, F105W, F110W, F125W, F140W, F160W. Moreover, it is possible to shift the star's spectrum due to the Doppler effect according to its line of sight velocity or due to the redshift of the entire GC. For the purpose of this paper, we did not apply any shift in the calculation. In this proof of concept study, we did not consider any source of absorption or reddening in our calculation.

The core radius (for which we mean the observational core radius) and the central surface brightness of each snapshot has been determined by applying a fit to the cumulative luminosity distribution (obtained from the snapshot) with the \cite{King1962} approximation ($L(r) = \pi \cdot R_c \cdot CSB \cdot ln (1 + (r/R_c)^2)$, with $L(r)$ being the luminosity at radius $r$, $R_c$ the core radius and $CSB$ the central surface brightness). The core properties are obtained as cumulative contribution of each star inside the core radius.

Finally, we applied a best fit (quadratic polynomial for Standard and BHS models; sum of two exponential for IMBH models\footnote{We used the best fit procedure present in Python scipy library, $scipy.optimize.curve\_fit$. }) to the luminosity weighted VDP, obtained from the infinite projection of the snapshot \citep{Mashchenko2005}. The velocity dispersion at different cluster radii has been determined by the value given by the best fit function at the desired radii (we consider the central velocity dispersion of the system to be the value obtained at $1\%$ light radius). A luminosity cutoff, for a star, of $5000\, L_\odot$\footnote{Value calculated from the maximum apparent magnitude, reddening and distance during observation of velocity dispersion in MWGCs \citep[e.g,][]{Carretta2009,Lane2011}. See \cite{Baumgardt2018} and citation therein for more informations.} has been applied to the determination of GCs' properties (total luminosity, properties at different light radii), in order to reduce the fluctuation due the presence in the system of only a few of very  luminous stars. The VDPs obtained from only luminous stars could be very noisy and strong fluctuations are expected due to the presence of luminous stars (in particular in the central part of IMBH models). Applying the fitting procedure to the velocity dispersion profile is the simplest approach to avoid those fluctuations while keeping the shape of the VDP.

\subsection{Model selection} \label{subsec:model-selection}
\citet{Askar2017} showed that models from the MOCCA Survey Database are in relatively good agreement with the observational properties of Galactic GCs (\cite{Harris1996}, updated 2010). However, the final goal for our research is to compare our models with EGGCs. For this reason, we should consider a sub-sample of the database which would mimic the observational limits and realistic properties of EGGCs.

The first limitation we imposed, is to consider only models that have $L > 2 \cdot 10^4 L_\odot$ (or equivalently, the absolute $M_V$ magnitude, $M_V < -6.5$), in order to mimic the observational limit luminosity for distant EGGCs (a nominal value of $L > 2 \cdot 10^4 L_\odot$ was chosen, even if with HST observations is possible to go below this value). The sample of selected models actually translates to the number of objects at $12\,\,Gyr$ $N > 10^5$ and the number of initial objects $N_0 > 10^5$\footnote{For models with $95\%$ binary fraction many binaries are dissolved  during the very early cluster evolution, so both limits on the minimum number of object (initial and at $12\,\,Gyr$) are important.}.

The second limitation we imposed, is to restrict our analysis to models in which the fallback prescription \citep{Belczynski2002} was used. Indeed, different prescriptions have been proposed to match the observed mass ranges and spin properties of stellar BHs, since the  observations of gravitational waves with LIGO/Virgo. The most accepted scenario is the presence of mass fallback on the BHs during SN explosion \citep{Belczynski2017}. We decided to limit our sub-sample to models for which this prescription has been applied.

The strength and the importance of the tidal field (that is the ``external'' field of host galaxy) is determined by the ratio between the tidal radius $R_t$ and the half mass radius $R_h$: models with higher $R_t/R_h$ ratios experience less influence from the tidal field with respect to ones with smaller ratios. For high values, the system is deeply inside its tidal field (i.e., all stars are deep inside the GC potential well), that would mean that the system will be able to freely expand before feeling the action of the tidal field. The system in this case is usually called `tidally underfilling model'. Instead, when the system fills its Roche lobe entirely, the importance of the tidal field in the evolution of the system is very strong (from the very beginning), substantially increasing the number of escaped stars. This system is usually called `tidally filling model'. So, in the case of a strong tidal field, and for a model that is tidally filling, the escape rate could be strong enough to dissolve the system in a time smaller than the Hubble time. \cite{Marks2012} find a weak relation between the half mass radius and the mass of newly formed star clusters ($R_h (pc) \propto (M/M_\odot)^{0.13}$), implying that at early stages, the clusters are very dense and strongly tidally underfilling. Recently, many theoretical works were published \citep{Marks2012,Kruijssen2014,Wang2020} supporting the idea that initially clusters are born very concentrated and deeply tidally underfilling. As it will be discussed in the next section, we will consider only systems that were initially tidally underfilling.

\section{Results} \label{sec:Results}
\subsection{Comparison with observed Galactic Globular Clusters } \label{sub-sec:MWGCs}
Firstly, we want to show that our selected sub-samples are in agreement with observations of MWGCs and their properties. This would ensure us that our model selection is able to represent observed GCs, and we can apply it also to the EGGCs. We assume that MWGCS are similar to EGGCs and the formation scenario for old GCs is similar.

\begin{figure}
    \centering
    \begin{subfigure}{0.5\textwidth}
       \centering
        \includegraphics[width=\linewidth]{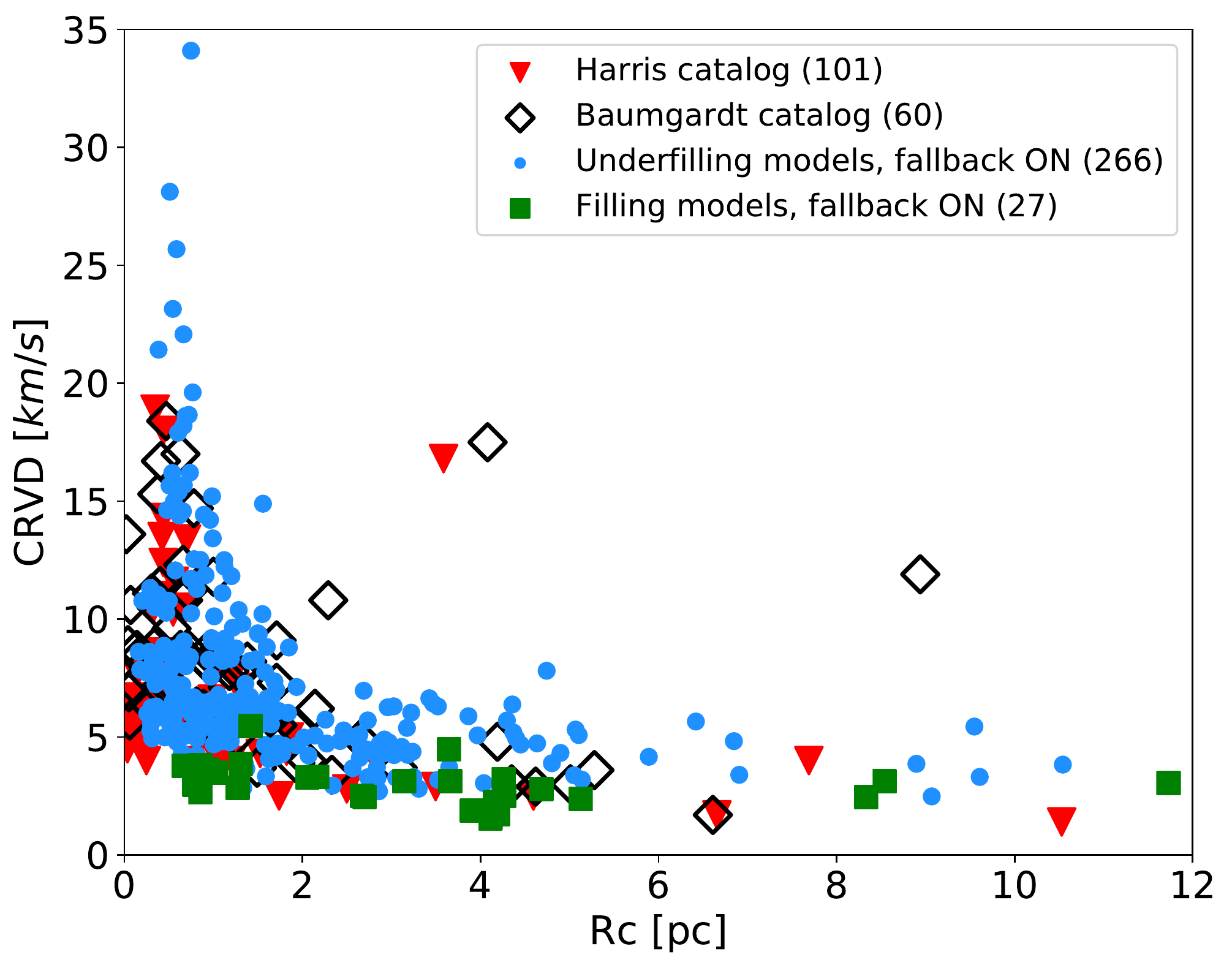}
    \end{subfigure}
    \begin{subfigure}{0.5\textwidth}
       \centering
        \includegraphics[width=\linewidth]{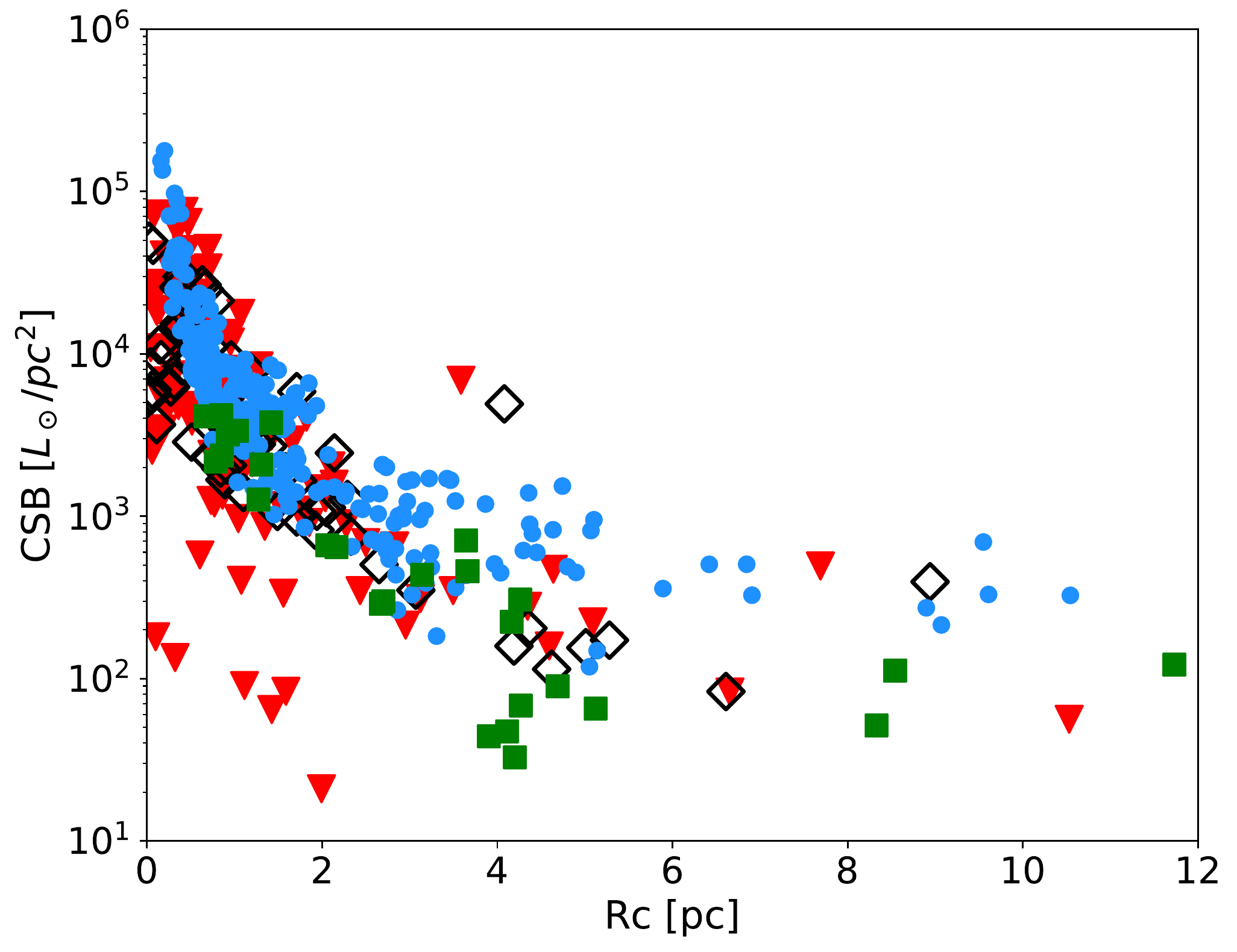}
    \end{subfigure}
     \caption{$R_c$  versus CRVD (top) and  CSB (bottom), for models at 12 Gyr with mass fallback prescription, tidally filling and underfilling models (blue circles and green squares, respectively) . Red triangles are the Galactic GCs from \citet[updated 2010]{Harris1996} catalog, and empty black diamonds from the Baumgardt catalog \citep{Baumgardt2018}. For those two catalogs, only GCs with $M_V < -6.5$ are plotted. In parenthesis the number of models for each catalog. }
     \label{Fig:csb-harris-comparison-fallb=1}
\end{figure}

In Fig. \ref{Fig:csb-harris-comparison-fallb=1} we compare the core radius ($R_c$) versus the central surface brightness (CSB) and central radial velocity dispersion (CRVD) respectively for our selected models with the \citet[updated 2010]{Harris1996} catalog and Baumgardt catalog \citep[private communication for central surface brigthness]{Baumgardt2018}. In \citet[updated 2010]{Harris1996} catalog, for some GCs the central structural parameters have not been observationally determined; for this reason, we selected only GCs for which those quantities are present in the catalogue and therefore only such MWGCs are presented in figures in this paper. Moreover, in order to compare our models with the two catalogs, we imposed  an absolute magnitude limit $M_V < -6.5$ 
(as imposed on our sub-sample) 
on the observational data. The total number of models in our selected sample is 266 for tidally underfilling models, meanwhile the \citet[updated 2010]{Harris1996} catalog has 101 and the Baumgardt catalog has 60. As one can see from those two figures, the tidally filling models cannot reproduce the high CSB and CRVD, as observed in Galactic GCs, but can actually match low CSB MWGCs, which are systems having low mass ($< 2 \cdot 10^5 \,M_\odot$) and low V absolute magnitude ($M_V \gtrsim -7$) in the considered sub-sample. Although we cannot exclude that some tidally filling models can reproduce properties of some low mass or close to disruption MWGCs, from the point of view of EGGCs such clusters will not be observable (or difficult to observe). Also the number of MWGCs which can be described by only tidally filling models is small, so if such models are not taken into account they will not spoil our statistic. For those reasons, we decided to exclude them from our sample and focus on underfilling models only.

Moreover, in Fig. \ref{Fig:csb-harris-comparison-fallb=1}, there are a few models with CRVD greater than $20\,\, km/s$, values that are not present in the observational catalog. Those models are connected with very massive IMBHs ($> 10^4 M_\odot$). Such high values for the IMBHs are obtained because in the simulation it is assumed $100\%$ accretion onto the BH; this is too optimistic of anassumption, so the real masses of IMBHs should be smaller than obtained in the MOCCA simulations. We decided to keep such models in our sample to show the properties of GCs which harbor such massive IMBHs. Since the number of such systems is small (only 6), this decision should not strongly influence our statistics.

\begin{figure*}
    \centering
    \begin{subfigure}{0.33\textwidth}
       \centering
        \includegraphics[width=\linewidth]{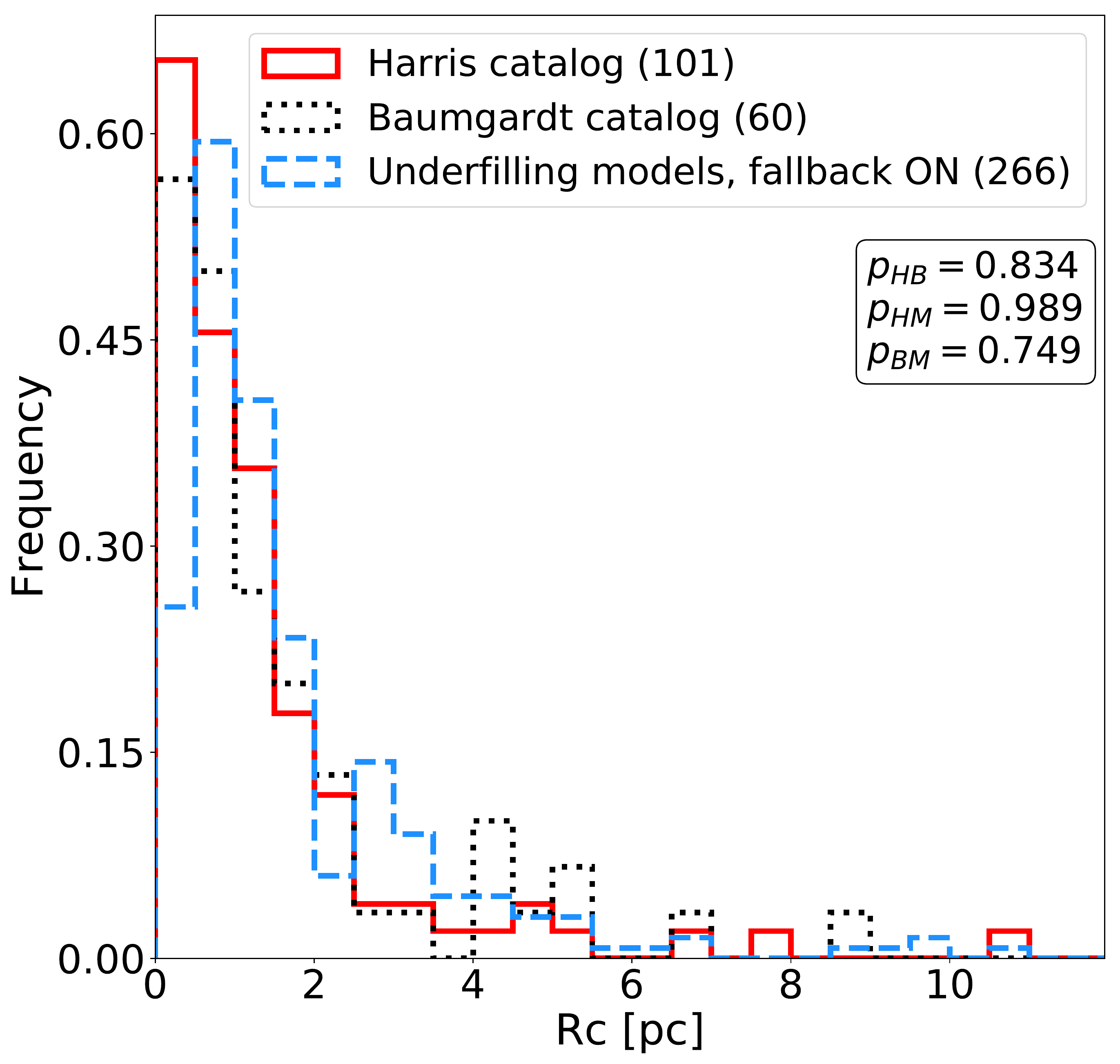}
    \end{subfigure}
    \begin{subfigure}{.33\textwidth}
       \centering
        \includegraphics[width=\linewidth]{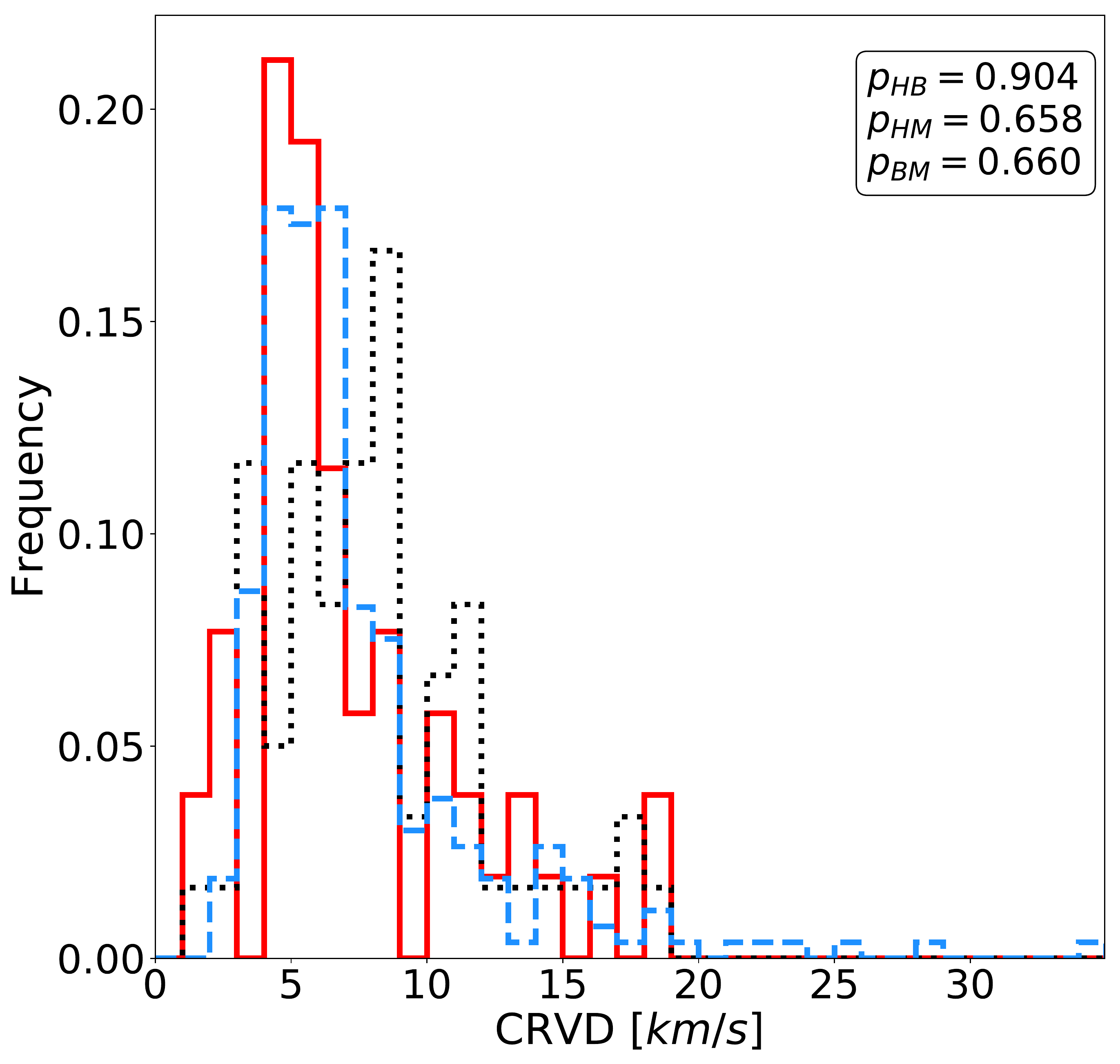}
    \end{subfigure}
    \begin{subfigure}{.33\textwidth}
       \centering
        \includegraphics[width=\linewidth]{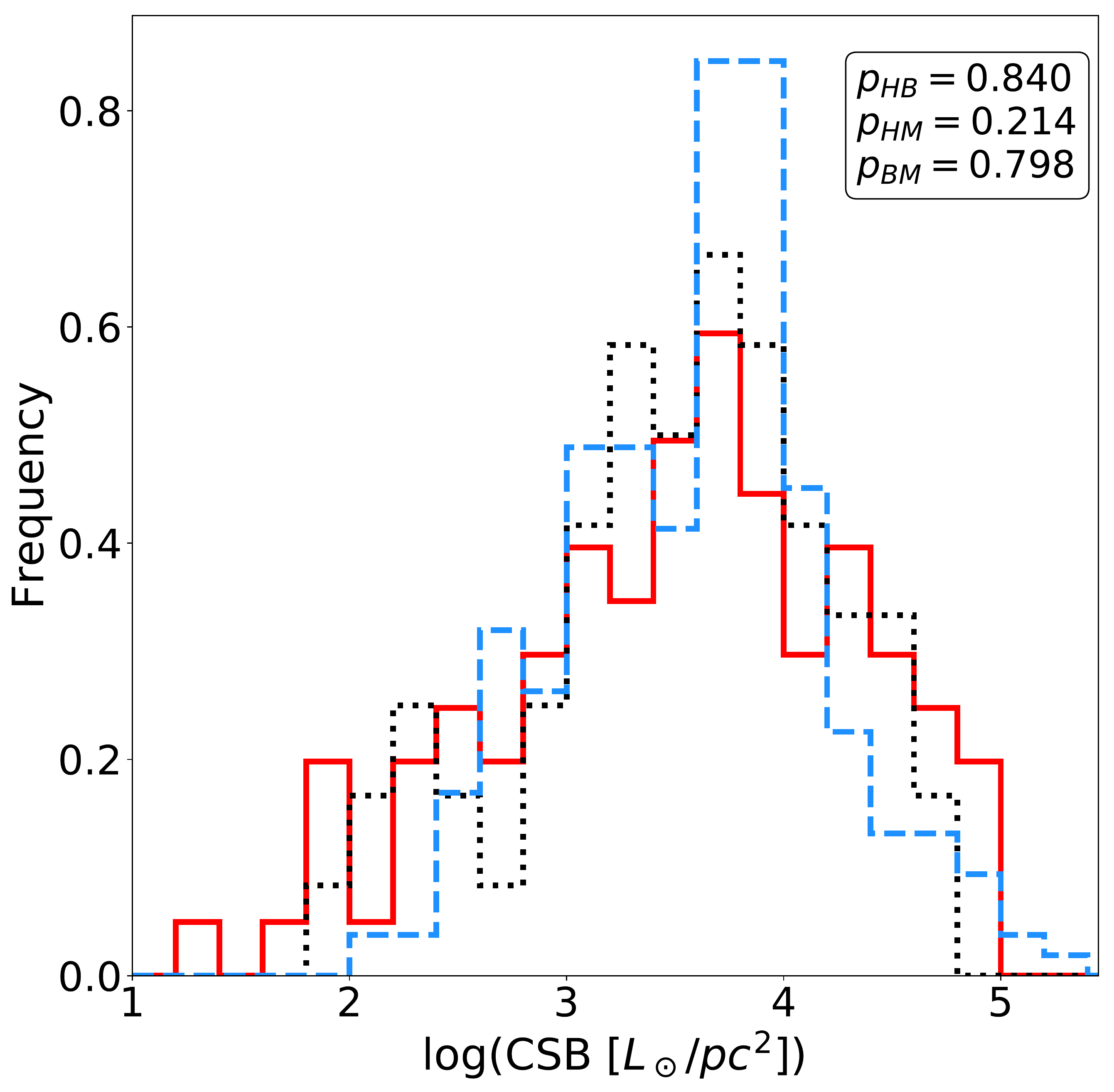}
    \end{subfigure}
\caption{$R_c$  (left), CRVD (centre) and CSB (right) histograms for our selected models (blue dashed), the observation of Galactic GCs from \citet[updated 2010]{Harris1996} catalog (red solid) and  the Baumgardt catalog \citep{Baumgardt2018} (black dotted). In parenthesis the number of models for each catalog. The histograms have been normalized so that the area under the histogram will sum to 1.0. The bin sizes are 0.5 ($R_c$ ), 0.7 (CRVD) and 0.2 (CSB) respectively. The panels on the right show the best p-value results from a Kolgomorov-Smirnov test (see text for more details). \textbf{Legend: HB}: Harris - Baumgardt catalogues comparison;  \textbf{HM}: Harris - MOCCA catalogues comparison;  \textbf{HB}: Baumgardt - MOCCA catalogues comparison.}
\label{Fig:hist-comparison}
\end{figure*}

 The distribution of the central parameters ($R_c$, CSB and CRVD) for the two catalogs and our selected models are shown in Fig. \ref{Fig:hist-comparison}. In order to verify that such models are statistically in agreement with the two catalogs, we applied a two-sample Kolgomorov-Smirnov test (KS test) to those distributions. The KS test is used to compare two sample, quantifying the distance between the cumulative distribution functions, in order to verify the null hypothesis that the two samples are drawn from the same distribution. However, it is also possible to apply the alternative hypothesis, according to which the cumulative distribution of one sample is ``less'' or ``greater'' than the cumulative distribution of the other sample. In KS test terminology, a cumulative distribution that is ``greater'' than one other means that its mean and median will be smaller than the mean and median of the other distribution (vice-versa for ``smaller''). In a few words, applying this alternative hypothesis means that the two distributions have the same shape, but the mean values are shifted, one with respect to the other. We applied also the alternative hypothesis (``less'' or ``greater'' ) to our sample, and a threshold value of $p \ge 0.17$, meaning a significance of $2 \sigma$ (instead, $p \ge 0.8$ would mean a significance of $3 \sigma$). Due to observational and systematic errors, it is possible that some values could be over- or under-estimated in the observed samples: a shift between the observed and our distributions may be expected, so the alternative hypothesis should be tested too. The p-values for different hypotheses (alternative and null) and for the three parameters are showed in Table \ref{Table:KStest-results}. The best p-values obtained are plotted in the upper-left boxes in  Fig. \ref{Fig:hist-comparison}. The results of the KS test show that our sample has a similar distribution to the two observational catalogues, with some differences: the $R_c$ of the two observational catalogues has a smaller mean radius compared to our sample; the Harris catalogue and our sample show a similar mean value for CSB, meanwhile our sample has a smaller mean CRVD when compared to Baumgardt catalog. In Table \ref{Table:KStest-results}, we also show the results of the KS test between the two observational catalogs. It is possible to see that the \citet[updated 2010]{Harris1996} catalog has smaller mean values compared to the \citep{Baumgardt2018} catalog. Considering that the two catalogs are based on the same observational data, but two different approaches have been used (in the \citet[updated 2010]{Harris1996} catalog, a fit of a \cite{King1966} profile has been applied to the observations, meanwhile in the \cite{Baumgardt2018} catalog a fit with N-Body simulation has been applied), that differ from our method, some systematic shifts in the distribution (between the three samples) are expected. We should remember that in the \citet[updated 2010]{Harris1996} catalog, for some collapsed GCs, it was arbitrarily assumed that the concentration parameter is equal to 2.5. Overall, those results show that our sample well represents the observed Galactic GCs; for this reason, we can assume that this will be true also for EGGCs.

\begin{table*}
    \centering
    \begin{tabular}{ c@{\hskip 0.5in} ccc ccc ccc}
    \hline
        & \multicolumn{3}{c}{$R_c$} & \multicolumn{3}{c}{CRVD } & \multicolumn{3}{c}{CSB}    \\
        \hline
        Hypothesis & HB & HM & BM  & HB & HM & BM  & HB & HM & BM \\
        \hline
        Less & \textbf{0.834} & \textbf{0.989} & \textbf{0.749} & \textbf{0.904} & \textbf{0.658} & 0.002 & 0.465 & 0.191 & 0.645 \\
        Two-sided &0.735 & 0.002 & 0.082 & 0.048 &0.371 & 0.004 & \textbf{0.840} & \textbf{0.214} & \textbf{0.798} \\
        Greater & 0.390 & 0.001 & 0.041 & 0.002 & 0.186 & \textbf{0.660} & 0.622 & 0.107 & 0.433\\
    \hline
    \end{tabular}
    \caption{Results of p-values from the Kolgomorov-Smirnov test, for different hypotheses and for $R_c$  (left), CRVD (centre) and CSB (right). The best p-values are marked in bold.  \textbf{Legend: HB}: Harris - Baumgardt catalogues comparison;  \textbf{HM}: Harris - MOCCA catalogues comparison;  \textbf{HB}: Baumgardt - MOCCA catalogues comparison.}
    \label{Table:KStest-results}
\end{table*}

\subsection{Dynamical model selections} \label{subsec:dyn-model-selection}

In the standard picture of the dynamical evolution of GCs, a GC would undergo a core collapse, 
unless a central source of energy can support the energy demand of the system. This supply is generally related to the energy released during the interaction of stars and binaries in the centre. After the early phase of SN explosions, some BHs can remain in the system (if the natal SN kick is not too high or if the gravitational potential is really deep) and quickly segregate in the center of the cluster, forming the so-called BH subsystem (BHS). This subsystem is not entirely decoupled from the rest of the GC and its evolution is governed by the energy demands of the host GC \citep{BreenandHeggie2013a,BreenandHeggie2013b,Giersz2019}. However, if the initial central density of the system is extremely high ($ >  10^8 M_\odot/pc^3$), the dynamical collision between massive BHs and runaway mergers of main sequence stars and BHs in the central region increase strongly, leading to the formation of an IMBH in the first phase of the cluster evolution (time smaller or roughly 1 Gyr; Fast scenario, \cite{Giersz2015}). If the first condition is not met, but not all BHs have been expelled from the system and only a few BHs are left in the system at the time of around of the cluster core collapse, an IMBH can be still formed via multiple mergers or collisions of BHs and other objects during dynamical interactions (Slow scenario, \cite{Giersz2015}). The dynamical evolution and properties of the GCs strongly depends on the presence of an IMBH (deep gravitational potential, kicking out all the massive BHs), or of a BHS, or the absence of both.

We dived our selected sample in three dynamical sub-samples, according to the following:
\begin{itemize}
    \item if an IMBH (BH with mass > 500 $M_\odot$) is present in the system, it has been classified as \textbf{IMBH model}; 
    \item if the number of BH ($N_{BH}$) present in the system is $\ge 50$, it has been classified as \textbf{BHS model}; if $20 < N_{BH} < 50$, we checked if the system is not experiencing the core collapse: if the system is in balanced evolution \citep{BreenandHeggie2013a,BreenandHeggie2013b}, it has been classified as \textbf{BHS model};
    \item if none of the previous conditions has been satisfied, the system has been classified as \textbf{Standard model}.
\end{itemize}

In the case of systems with $20 < N_{BH} < 50$, we fit a third order polynomial to the evolution of the $10\%$ Lagrangian radius of the system; if the time derivative of this quantity at 12 Gyr and 13 Gyr is negative and less than $- 2.5 \cdot 10^{-3} \, pc/Gyr$, the system is considered in collapse, and classified as a Standard model, otherwise as a BHS model. The limit on the number of BHs was chosen by analysing many models by eye: we checked that systems having a number of BH  greater than 50 were dynamically in balanced evolution, meanwhile this is not always true for a small number of BHs. The choice of a minimum mass of $500 \,M_\odot$ for a BH to be classified as IMBHs follows from the fact that for smaller masses the IMBH will still substantially move around in the central parts of the system (see \cite{Giersz2015}), so their influence on the system structure will be smeared out and also the central structure will still be similar to that of a recently collapsed cluster.

The numbers of models for each sub-model category are: IMBH - 104; BHS - 93; Standard - 69. We would like to strongly emphasise that the relative number of models with different evolution types depends on the initial conditions chosen for the MOCCA simulations and should not be taken as a real number which can be confirmed by observations.

\subsection{Comparison with previous works} \label{sub-sec:comp-prev-works}
\begin{figure}
    \centering
    \includegraphics[width=0.5\textwidth]{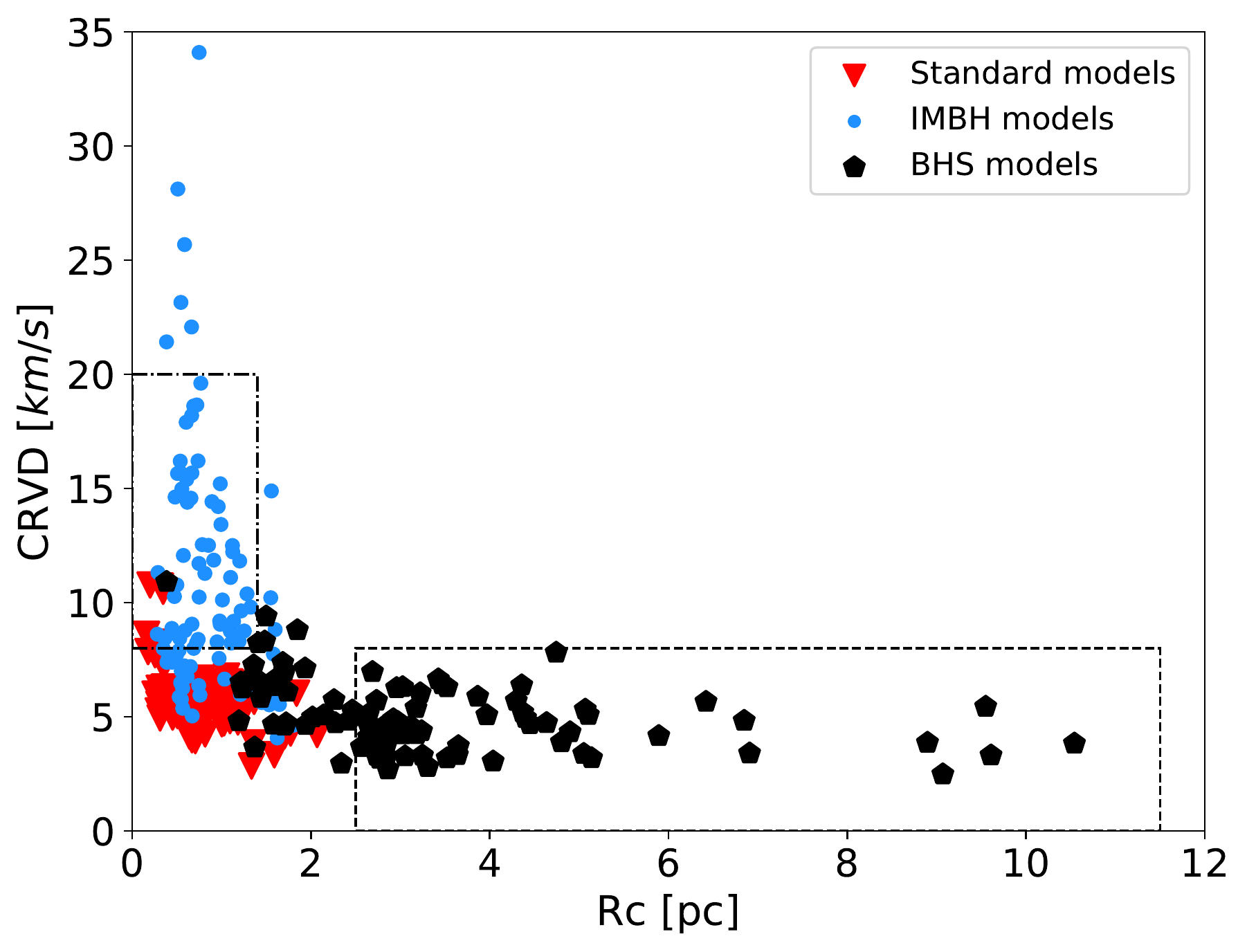}
    \hfill
    \includegraphics[width=0.5\textwidth]{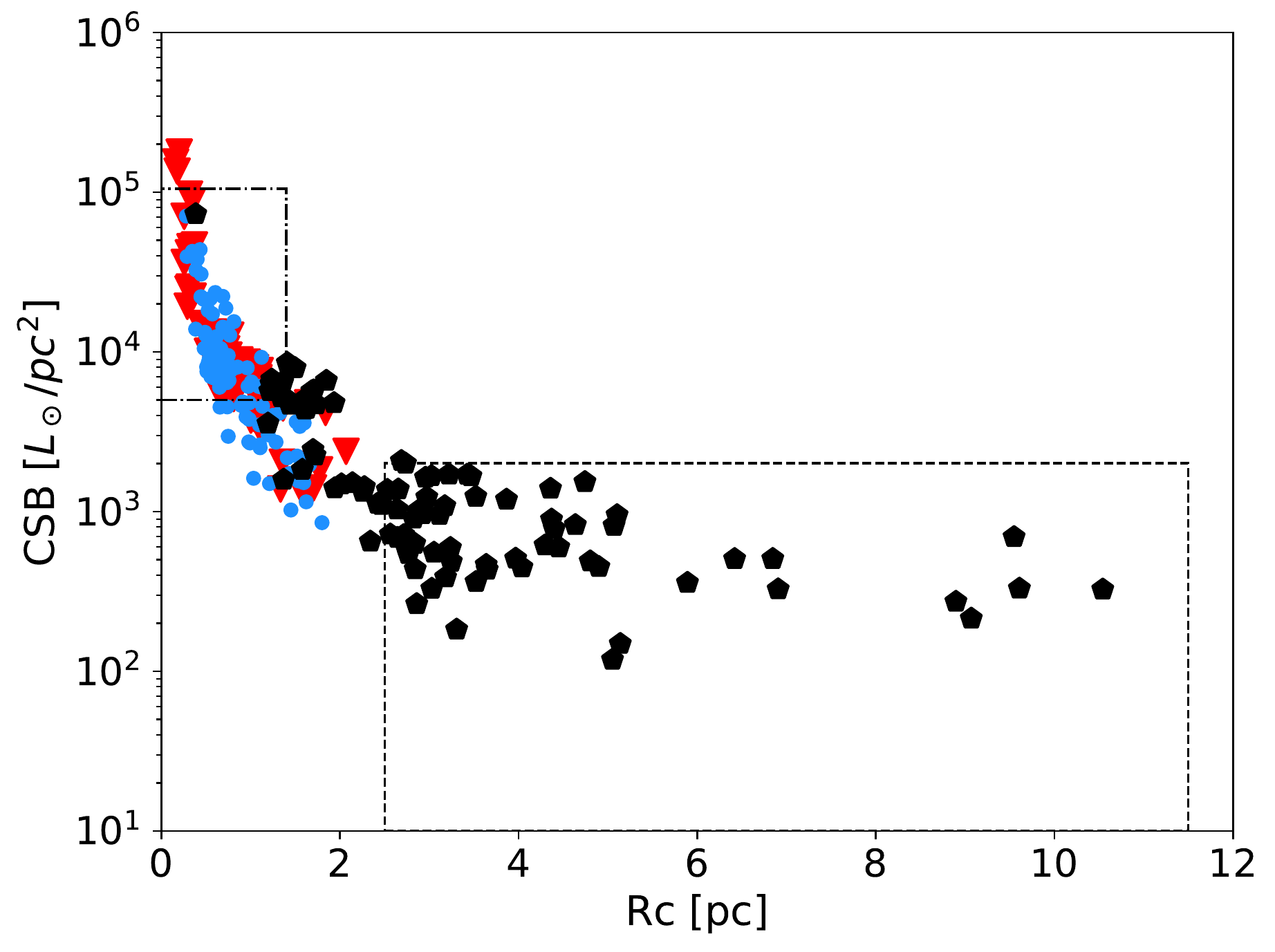}
    \caption{$R_c$  versus CRVD (top) and  CSB (below), for the three different dynamical models (red triangles for Standard, blue circles for IMBH and black pentagons for BHS, respectively). The dashed lines represent the cut applied for ``large core radii'' clusters ($R_c \geq 2.5\,pc$, $CRVD \leq 7.5\,km/s$, $CSB < 2\cdot10^3$), meanwhile the dashed-dotted line represent the cut applied for ``small core radii'' clusters ($R_c \leq 1.4 \,pc$, $CRVD\geq 8.0\,km/s$, $5\cdot 10^3 < CSB < 10^5$). }
    \label{Fig:dyn-rc-csb-sigma}
\end{figure}

In Fig \ref{Fig:dyn-rc-csb-sigma}, we show the position of such sub-models in the $R_c$  - CSB plane and in the $R_c$  - CRVD plane. As expected, the BHS models have (on average) a large $R_c$  value ($\gtrsim 2.0 \,pc$) and relatively low CRVD ($ \lesssim 7.5 \, km/s$) and CSB ($ \lesssim 10^4 \, L_\odot/pc^2$), meanwhile systems with an IMBH show a small core ($< 2.0 \,pc$) and high values for the central parameters (CRVD can reach values of $> 12.0 km/s$ and CSB can reach values of $> 10^4 L_\odot/pc^2$). However, both sub-models overlap with the Standard models in those two planes. This complicates the proper distinction between those models from an observational point of view: one can clearly see that such pairs of structural parameters are not enough for this purpose.

We applied some boundary conditions for observational cluster properties, in order to define two regions where only (or mostly) IMBH and BHS models are present, named ``small core radii clusters'' and ``large core radii clusters'' respectively. The ``small core radii clusters'' (IMBH) models have:
\begin{itemize}
    \item $R_c \leq 1.4 \,pc$;
    \item $8.0 < CRVD <  20.0\,km/s$;
    \item $5\cdot 10^3 < CSB < 10^5 \,L_\odot/pc^2$,
\end{itemize}

meanwhile ``large core radii clusters'' (BHS) models have:
\begin{itemize}
    \item $R_c \geq 2.5 \,pc$;
    \item $CRVD \leq 7.5\,km/s$;
    \item $CSB < 2\cdot10^3\,L_\odot/pc^2$.
\end{itemize}

With this selection criteria, the total number of ``large core radii clusters'' model is 59, of which all are BHS models, with no IMBH and Standard models; on the other hand,  the total number of ``small core radii clusters'' model is 49, of which 43 are IMBH models, 5 Standard models and only 1 BHS model.

In table \ref{Table:strongly-concentrated} and \ref{Table:weakly-concentrated}, we show the list of observed GCs that satisfied the conditions for  ``small core radii'' and ``large core radii'' models respectively, from Harris and Baumgardt catalogs. We consider only GCs for which all properties are available (that are $R_c$, CRVD and CSB), since in the Harris catalog not all system has a CRVD value and in the Baumgardt catalog the CSB is not given for all models (private communication). We have considered only those clusters that have been selected in both catalogs (Harris and Baumgardt).

\begin{figure*}
\begin{subfigure}{\columnwidth}
    \centering
    \includegraphics[width=\textwidth]{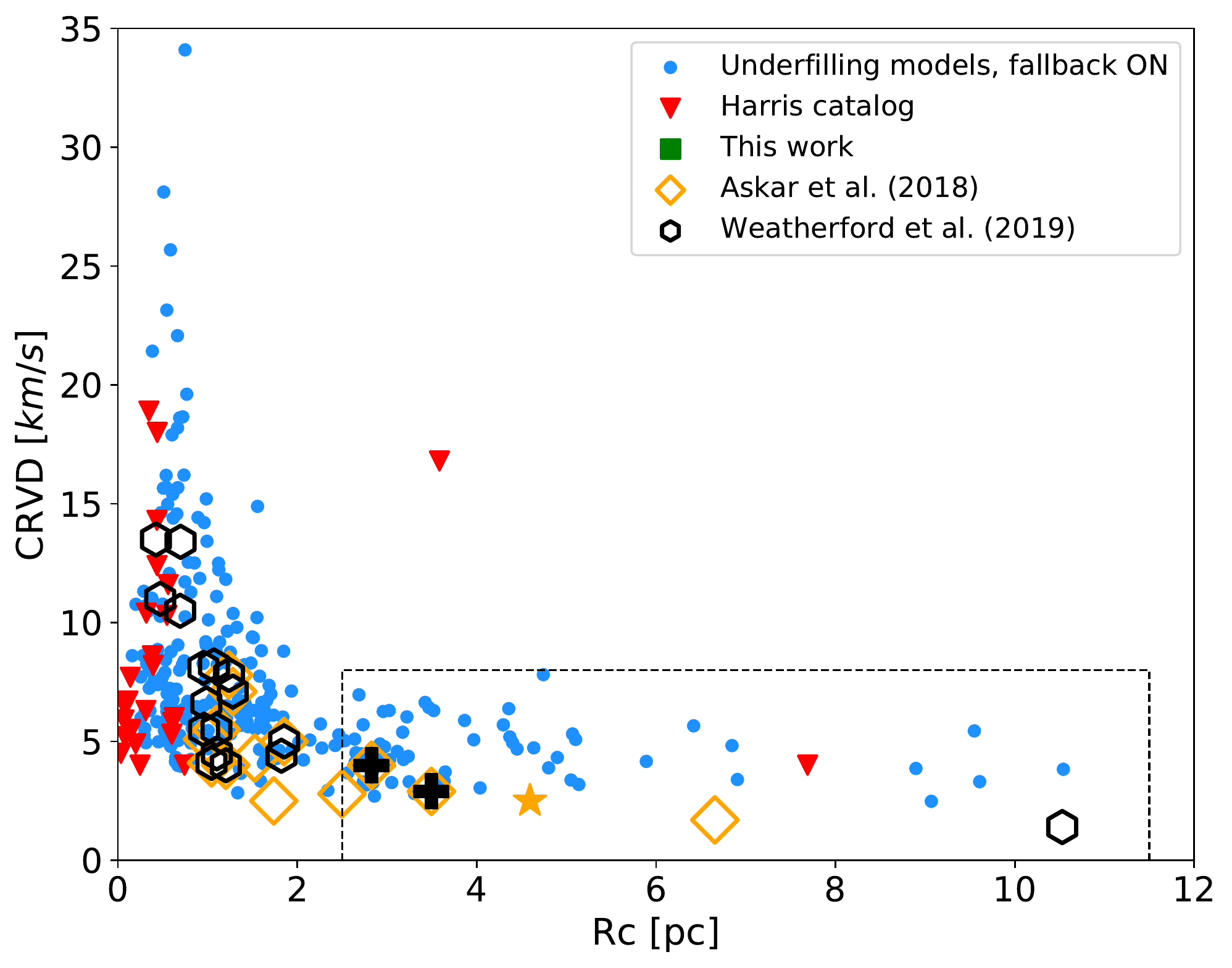}
    \hfill
    \includegraphics[width=\textwidth]{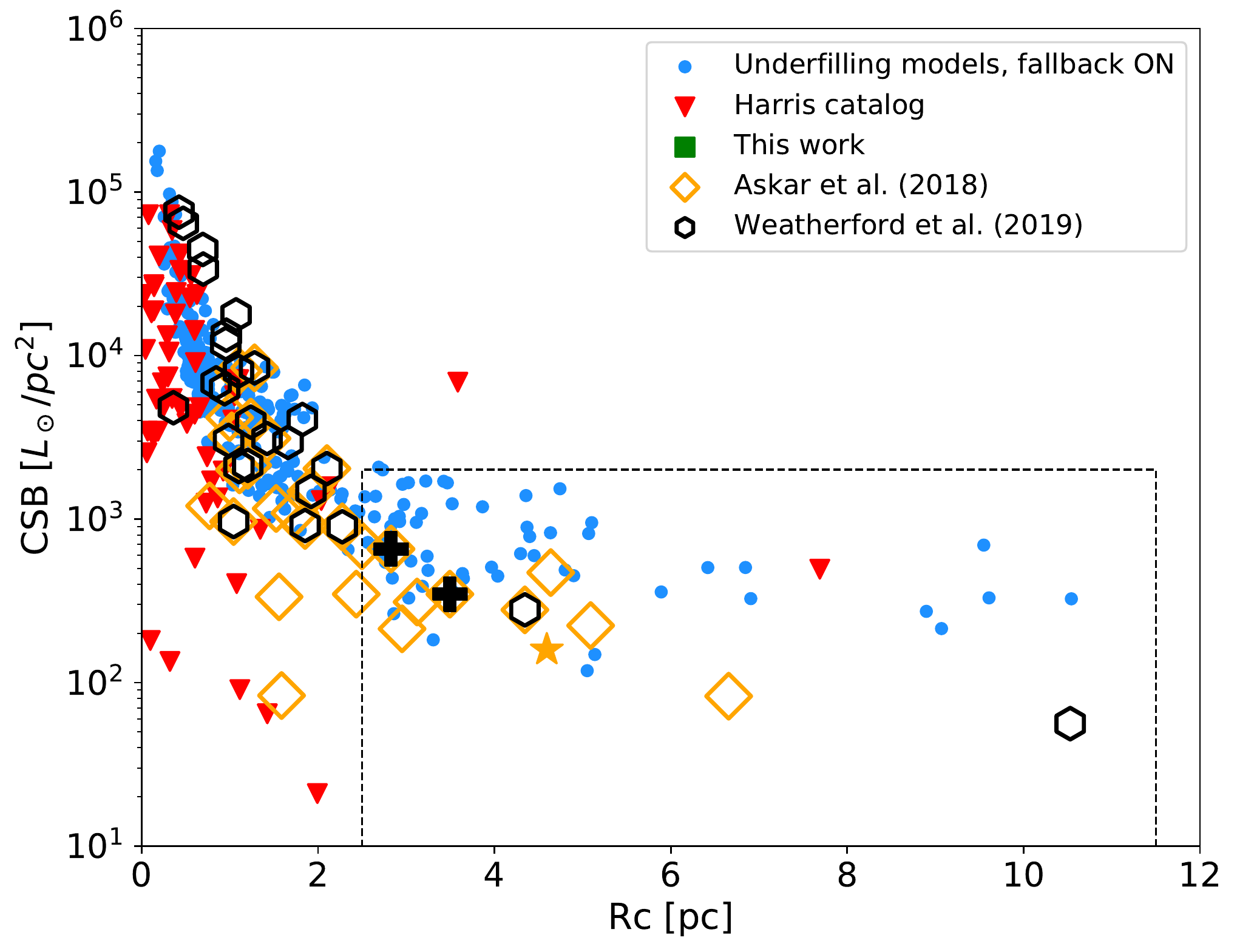}
  \end{subfigure}
  \begin{subfigure}{\columnwidth}
    \centering
    \includegraphics[width=\textwidth]{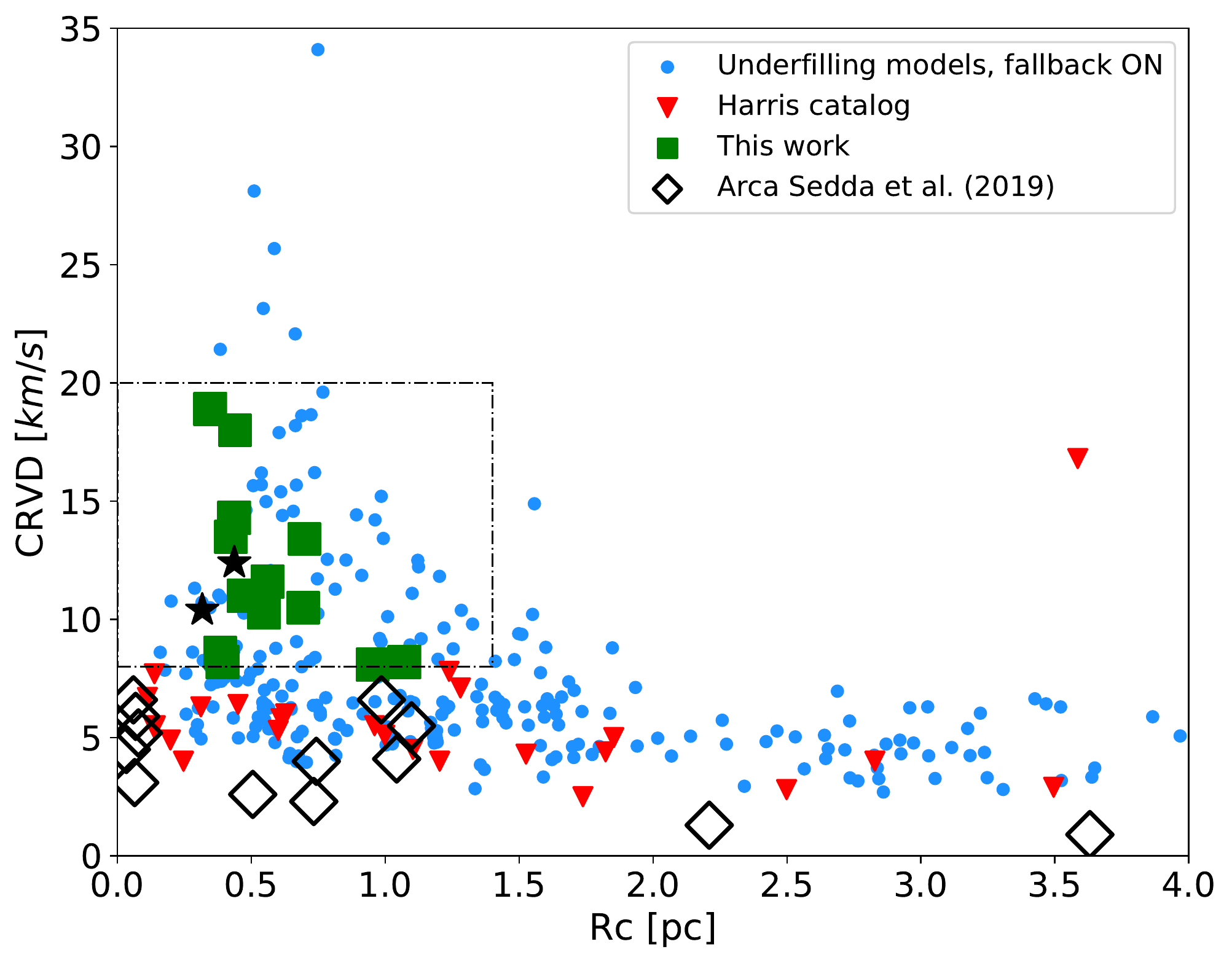}
    \hfill
    \includegraphics[width=\textwidth]{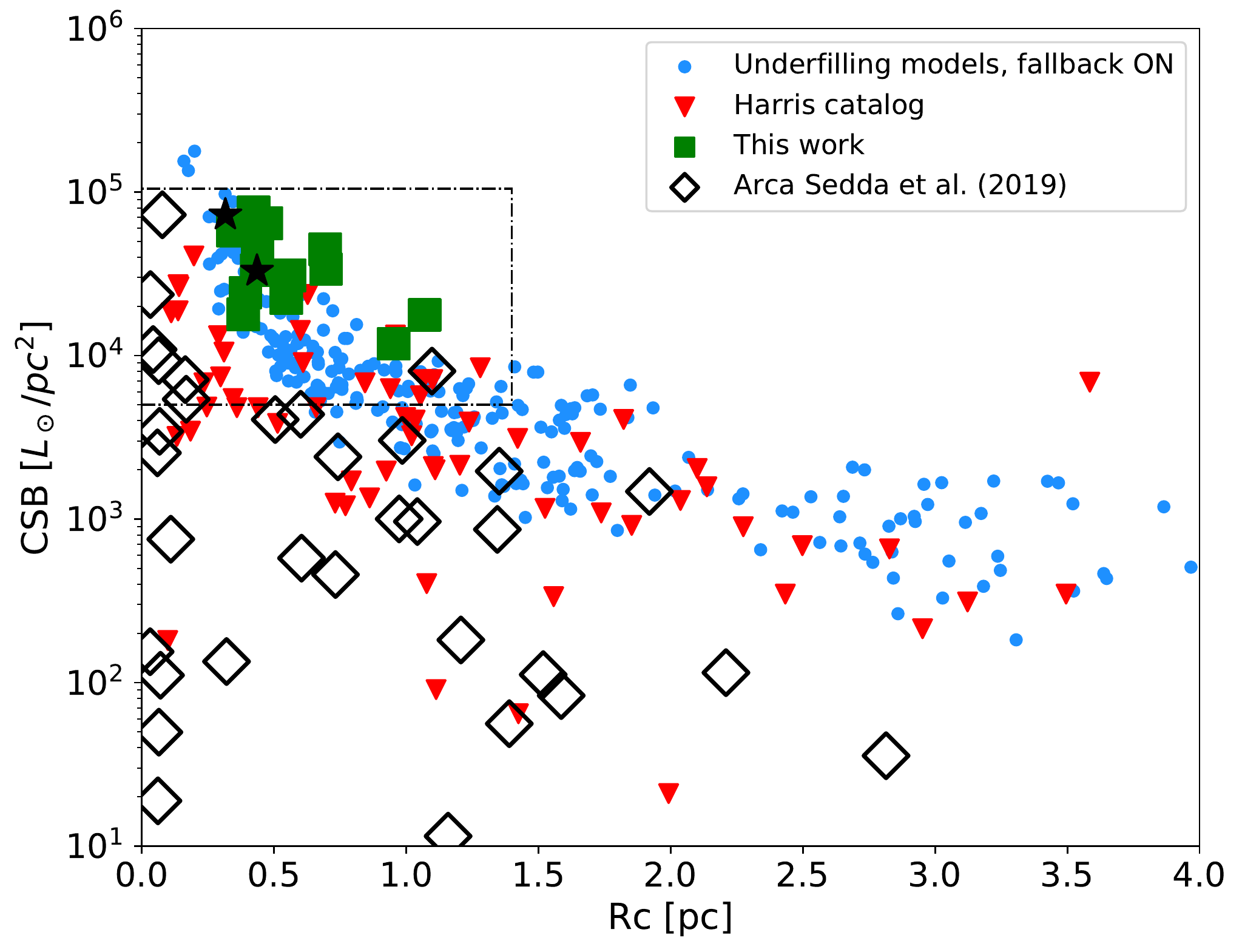}
  \end{subfigure}
  \caption{Same as Fig. \ref{Fig:dyn-rc-csb-sigma}, on the left for ``large core radii'' models and on the right for ``small core radii'' models, respectively. The models reported in this paper are shown with green squares. Models from \citet{Askar2018} and \citet{Weatherford2019} are shown in empty orange diamonds and in black hexagons, respectively (on the left) and models from \citet{ArcaSedda2019} are shown in empty black diamonds (on the right). Models that have been reported also in those papers are shown with stars (orange from \citet{Askar2018} and black from  \citet{ArcaSedda2019}, respectively) and in black crosses from \citet{Weatherford2019}. We expanded the region between $0.0 < R_c < 4.0\,pc$ for the figures in the right, in order to better distinguish between different points.}
    \label{Fig:comparison}
\end{figure*}

In Fig. \ref{Fig:comparison}, we compared our ``large core radii'' models ( which correspond to BHS models), with the list of GCs reported in \cite{Askar2018} which could contain a BHS, and the list reported in \cite{Weatherford2019}, considering only GCs that retain a number of BH $N_{BH} > 50$. Meanwhile in \cite{Askar2018} a shortlist of 29 Galactic GCs have been reported and the number of GCs shortlisted in \cite{Weatherford2019} with  $N_{BH} > 50$ is of 28 Galactic GCs, in this paper only three have been reported. However, some of the GCs listed in  \cite{Weatherford2019} with  $N_{BH} > 50$ are actually in the ``small core radii'' models region. This is not completely surprising, since the author did not consider a correlation between the CSB and number of BH inside the GC. Finally, we compared our ``small core radii'' clusters (that correspond mostly to IMBH), with the list of GCs reported in \cite{ArcaSedda2019}. In their work, 35 observed Galactic GCs are likely to harbour an IMBH, meanwhile in our work only 15 do. Even if the number of reported GCs are much smaller in this paper compared to previous works, one can note that the actual number of observed MWGC that are inside our defined regions would increase (up to roughly 17 for  ``large core radii'' region and up to roughly 30 for ``small core radii'' ones), if we were not to restrict our consideration to only GCs which have all properties available. 

The source of observed large discrepancies will be broadly discussed in more detail in Section \ref{sec:Discussion}, where possible reasons will be provided. Overall, our model selection of ``small'' and ``large core radii cluters'' is in rough agreement with the results of previous work.

\begin{table*}
    \centering
    \begin{tabular}{ ccccc}
    \hline
    Name & $R_{c}\,\,(pc)$ & $R_{hl}\,\,(pc)$ & $CRVD \pm \delta_{CRVD} \,\,(km/s)$ & CSB ($L_\odot/pc^2$) \\
        \hline
    NGC104 &  0.47 & 4.15 &$11.0 \pm 0.3 $ & 64394.67 \\
    \textbf{NGC1851}  & \textbf{0.32} & \textbf{1.79} & \bm{$10.4\pm 0.5$} & \textbf{72585.49}\\
    NGC2808 & 0.70 & 2.23 &  $13.4 \pm  1.2 $ &  33794.80 \\
    \textbf{\textit{NGC5286}} & \textbf{\textit{0.95}} & \textbf{\textit{2.48}} & \textbf{\textit{8.1}} $\bm{\pm}$  \textbf{\textit{1.0}} & \textbf{\textit{11826.34}} \\ 
    NGC5824  & 0.56 & 4.20 & $11.6 \pm 0.5$ & 30821.20\\
    \textbf{NGC6093} & \textbf{0.43} & \textbf{1.77} & \bm{$12.4 \pm  0.6 $} &  \textbf{32873.81} \\
    NGC6266 & 0.43 & 1.82 & $14.3  \pm 0.4 $ &  33177.97 \\
    NGC6388 & 0.34 & 1.50 & $18.9 \pm  0.8  $ & 58190.21 \\
    NGC6541 & 0.39 & 2.31 & $8.2 \pm  2.1 $ &  24257.73 \\
    NGC6626 & 0.38 & 3.15 & $8.6 \pm  1.3 $ &  17899.87 \\
    NGC6864  & 0.55 & 2.80 & $10.3 \pm 1.5$ &22534.59\\
    NGC7078 & 0.42 & 3.02 & $13.5 \pm  0.9 $ &  75309.51 \\
    NGC7089 & 1.07 & 3.55 & $8.2  \pm 0.6  $ & 17735.77 \\
    \hline
    \end{tabular}
    \caption{Observational properties from the \citet[updated 2010]{Harris1996} catalog (name, $R_c$  and half-light radius in parsec, CRVD in $km/s$ and CSB in $L_\odot/pc^2$) for observed Galactic GC labelled as ``small core radii'' in this work, that are likely to host an IMBH. In bold the clusters that have been reported also in \citet{ArcaSedda2019}, in bold italic the clusters that have been reported also in \citet{Lutzgendorf2013}.}
    \label{Table:strongly-concentrated}
\end{table*}

\begin{table*}
    \centering
    \begin{tabular}{ ccccc}
    \hline
    Name  & $R_{c}\,\,(pc)$ &$R_{hl}\,\,(pc)$ & $CRVD \pm \delta_{CRVD} \,\,(km/s)$ & CSB ($L_\odot/pc^2$)  \\
        \hline
 \textbf{\textit{NGC288}}  & \textbf{\textit{3.49}} & \textbf{\textit{5.77}} & \textbf{\textit{2.9}} $\bm{\pm}$ \textbf{\textit{0.3}}  & \textbf{\textit{347.42}}\\
 \textbf{IC4499} & \textbf{4.59} & \textbf{9.35} & \bm{$2.5 \pm 0.5 $} & \textbf{158.80}\\
 \textbf{\textit{NGC6809} } & \textbf{\textit{2.83}} & \textbf{\textit{4.44}} & \textbf{\textit{4.0}} $\bm{\pm}$ \textbf{\textit{0.3}} & \textbf{\textit{655.92}}\\

    \hline
    \end{tabular}
    \caption{Observational properties from \citet[updated 2010]{Harris1996} catalog (name, $R_c$  and half-light radius in parsec, CRVD in $km/s$ and CSB in $L_\odot/pc^2$) for observed Galactic GC labelled as ``large core radii'' in this work, that are likely to host a BHS. In bold the clusters that have been reported also in \citet{Askar2018}, in bold italic the clusters that have been reported also in \citet{Weatherford2019}.}
    \label{Table:weakly-concentrated}
\end{table*}

\subsection{Color distribution} \label{subsec:color}
A well know property of observed Galactic and EGGCs is the bi-modality (or sometimes even multi-modality) in color distribution. Even if it is up for debate whether or not it is universal features of GCs, it seems to be observed in most of the GC populations around early-type galaxies. This feature has been correlated with a bi-modality (or multi-modality respectively) in metallicity.

\begin{figure}
    \centering
        \includegraphics[width=0.5\textwidth]{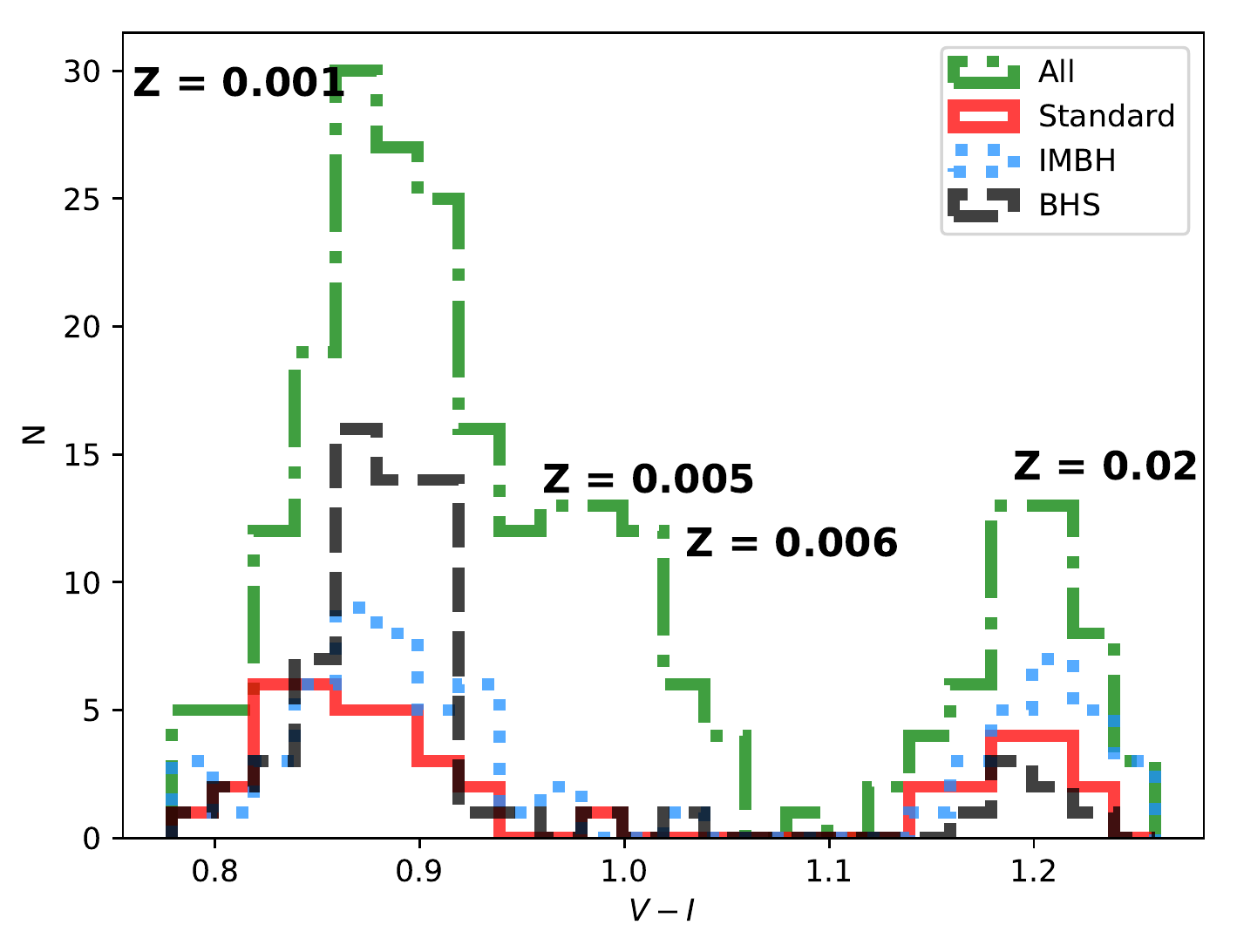}
    \caption{$V-I$ color distribution for models with solar metallicity $Z=0.02$ ($[Fe/H] = 0.0$) and sub-solar metallicities $Z=0.001$ ($[Fe/H] = -1.3$), $Z=0.005$ ($[Fe/H] = -0.55$), $Z=0.006$ ($[Fe/H] = -0.6$). The corresponding metallicity for each peak is written to the side. For $Z=0.02$ and $Z=0.001$ metallicities, the distribution for different dynamical models is also reported, meanwhile for $Z=0.005$, $Z=0.006$ metallicities only the total distributions have been reported. }
    \label{Fig:VI-dist}
\end{figure}

We tested, for fixed metallicities, the V-I color distribution for our sample of models. In Fig. \ref{Fig:VI-dist} we show the V-I color distribution for models with solar metallicity ($Z=0.02$, corresponding to $[Fe/H] = 0.0$; number of models: 45) and for subsolar metallicities ($Z=0.001$, $Z=0.005$, $Z=0.006$, corresponding to $[Fe/H] = -1.3$, $[Fe/H] = -0.55$, $[Fe/H] = -0.6$ respectively; number of models: 137, 49, 18, respectively). As can be seen from the figure, the distribution for each metallicity is uni-modal, but the spread in color can be substantial (of the order of $\sim 0.2\,\, dex$ for subsolar metallicity, color spread that would correspond to a spread in metallicity of $[Fe/H] \sim 0.6$\footnote{We used the calibration in \citet{Kissler-Patig1998}, as was done in \cite{Larsen2001}.}), in agreement with the observational distribution \cite{Larsen2001}. From this figure it is also possible to note some signs of a slightly different spread in the color, for different dynamical states, at fixed metallicity: this could imply that the internal dynamical evolution and state of GCs could influence the color distribution. Finally, the value of this spread is comparable to the spread due to a small difference in metallicities: the color spread for sub-metallicities  $Z=0.001$ and $Z=0.006$ is of the order  $\sim 0.2\,\, dex$. However, this is difficult to confirm, since the number of models is not big enough to be statically strong and further studies are needed. 

The importance of the spread in color distribution for models at fixed metallicity could be explained by the interplay of different initial conditions and different dynamical history. Initial concentration and initial binary fraction are the main properties that drive the dynamical evolution of the system (core collapse, disruption,..), since they mostly influence the interaction and collision rate of stars: lower or higher numbers of stars could  be removed, depending on the density and concentration of the system. The general fate of the system has been explained in Sec. \ref{subsec:dyn-model-selection}, which depends on the initial conditions. This could be even enhanced by the presence of an IMBH (and depending on the mass of the IMBH), by a strong tidal field, or during core collapse (in Standard models, for example).

\begin{figure}
    \centering
    \begin{subfigure}{0.5\textwidth}
       \centering
        \includegraphics[width=\linewidth]{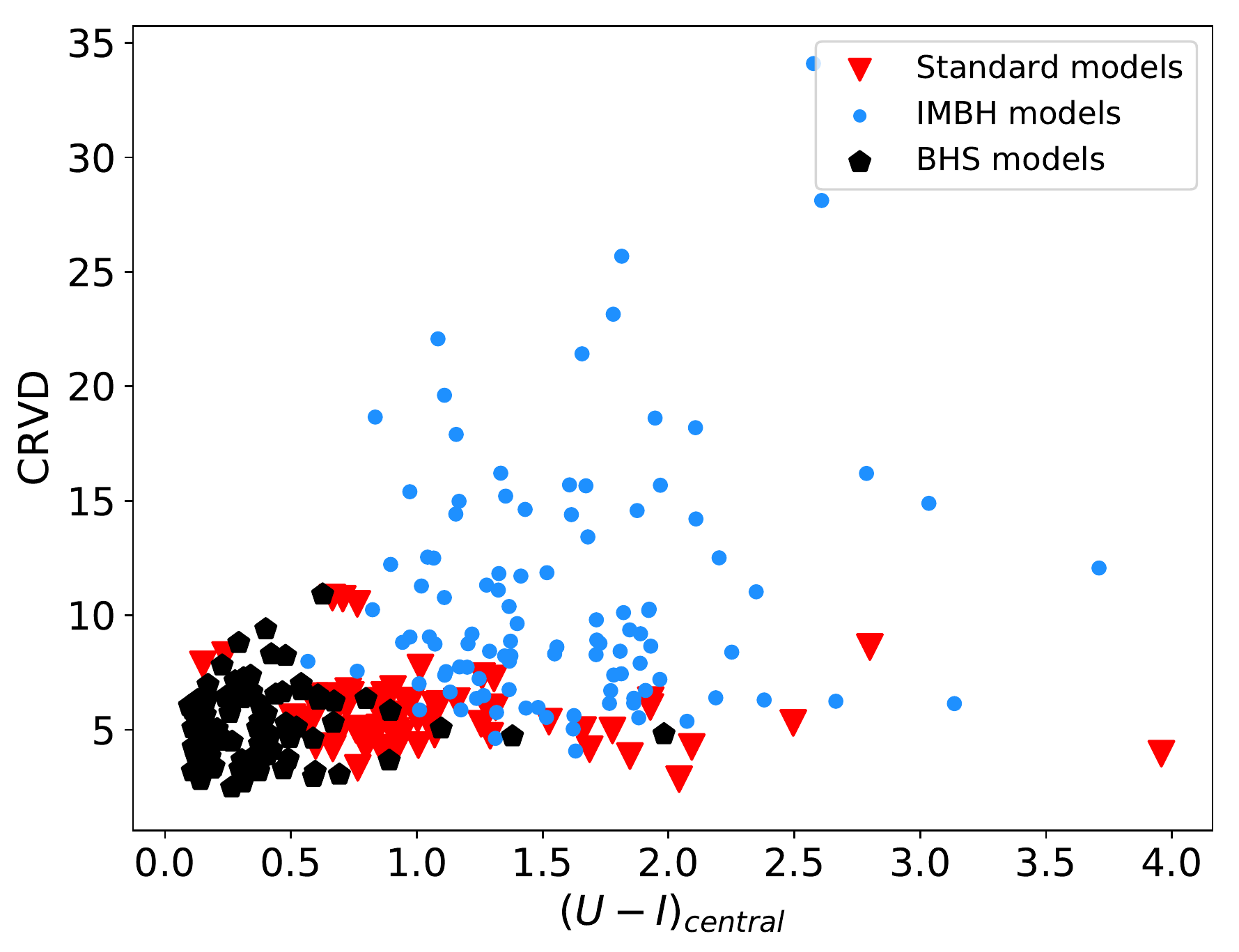}
    \end{subfigure}
    \begin{subfigure}{0.5\textwidth}
       \centering
        \includegraphics[width=\linewidth]{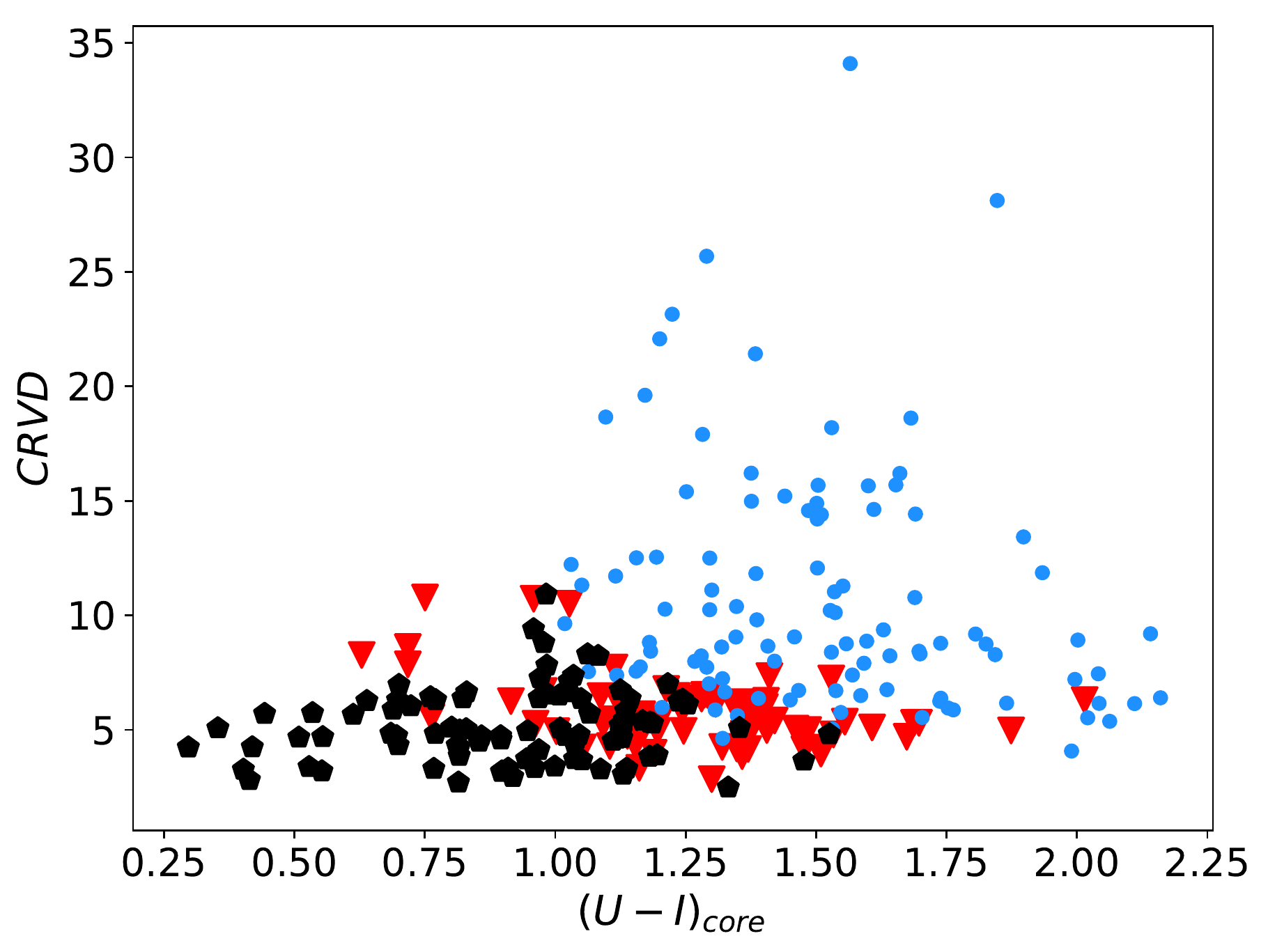}
    \end{subfigure}
    \begin{subfigure}{0.5\textwidth}
       \centering
        \includegraphics[width=\linewidth]{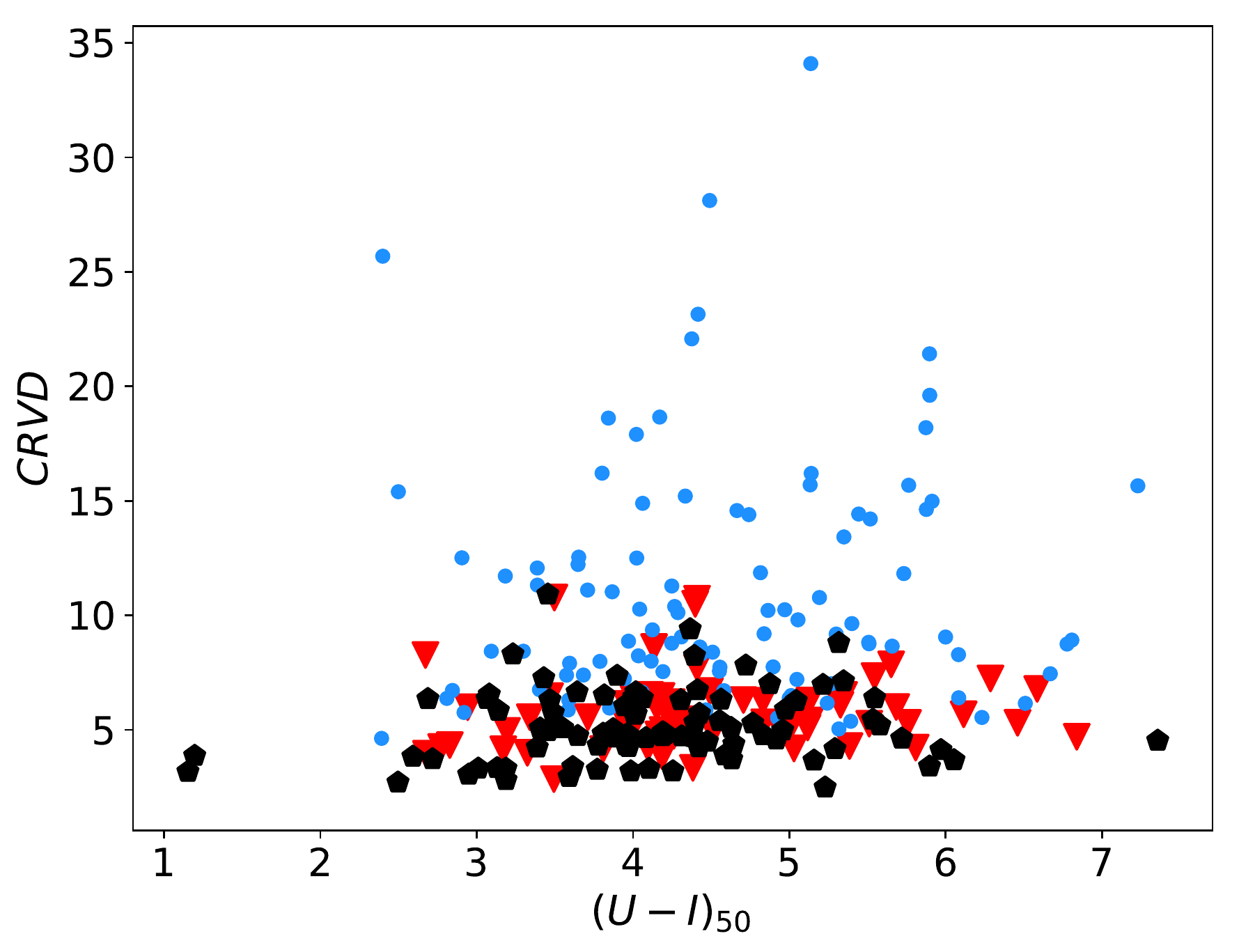}
    \end{subfigure}
     \caption{Central $U-I$ color (top), $U-I$ color at $R_c$  (middle) and at $R_{hl}$ (bottom) versus CRVD, respectively, for different dynamical models. The distinction among different dynamical models is the best for the central value color, and it gets worse for larger radii.}
     \label{Fig:color-gradient}
\end{figure}

In Fig. \ref{Fig:color-gradient}, we show the $U-I$ color at different radii (central, core and $R_{hl}$) versus CRVD, CRVD. It is possible to note that the models are barely indistinguishable when considering the color at $R_{hl}$, and the differentiation is getting better for smaller radii. However, it is not possible to distinguish properly the three dynamical models only considering two properties of the system, even if those are the central properties. In order to better separate the different dynamical models, more than two (central) properties of the system are needed.

\subsection{3D space parameters} \label{subsec:3D}
As was shown in the previous sections, considering only two global GCs' properties is not enough to divide the dynamical models into three different regions of the parameter space. In this section we will discuss how this is possible in 3D space.  To make the comparison with observations as easy as possible, we will try to use properties which can be in principle observed now or in future observation campaigns. To remind the reader, in this paper the central properties have been defined as the values obtained at $1\%$ light radius, meanwhile the core properties are obtained as the cumulative contribution of each star inside the $R_c$ .

\begin{figure*}
\begin{subfigure}{\columnwidth}
        \centering
        \includegraphics[width=\textwidth]{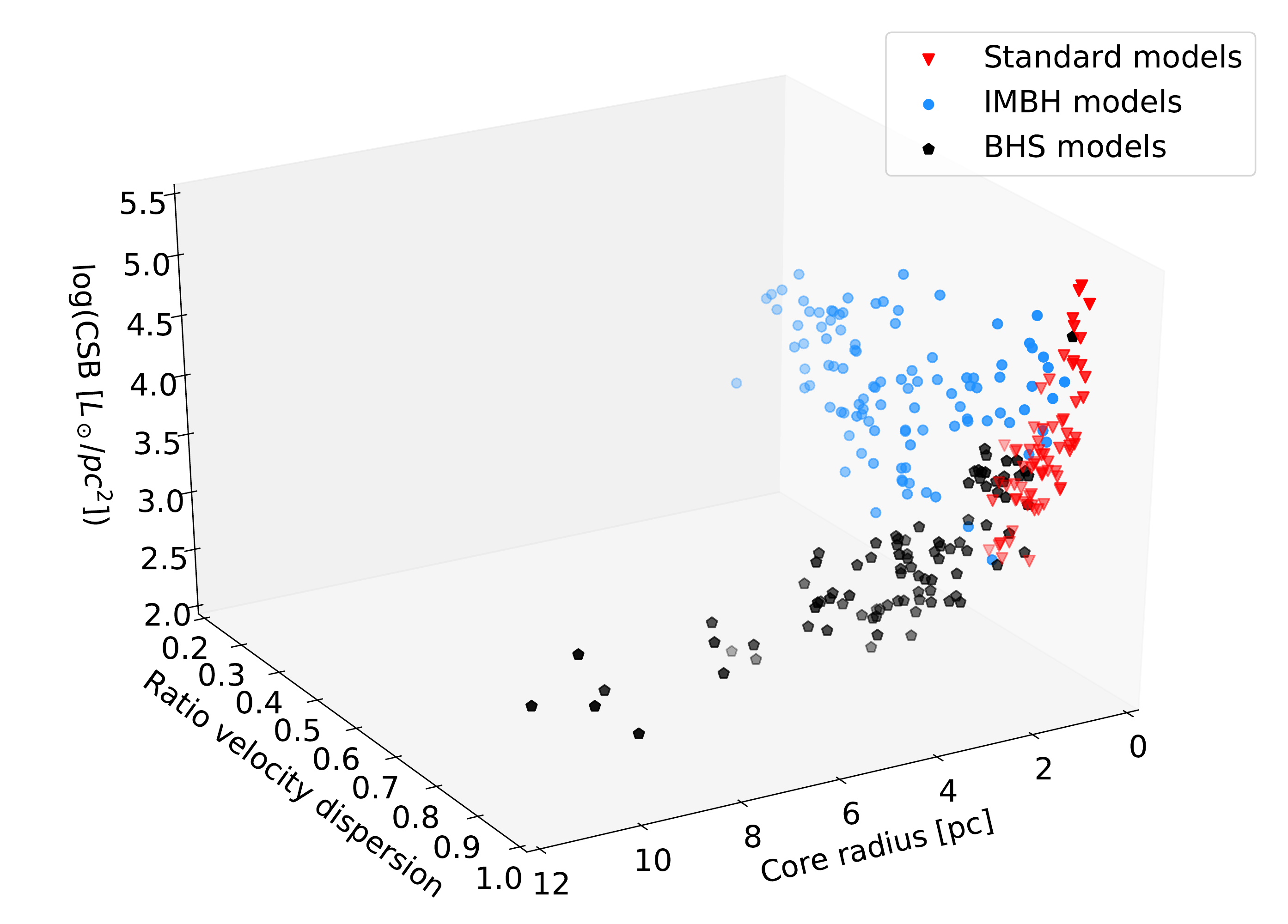}\\
        \hfill
        \vspace{10.5pt}
        \centering
        \includegraphics[width=\textwidth]{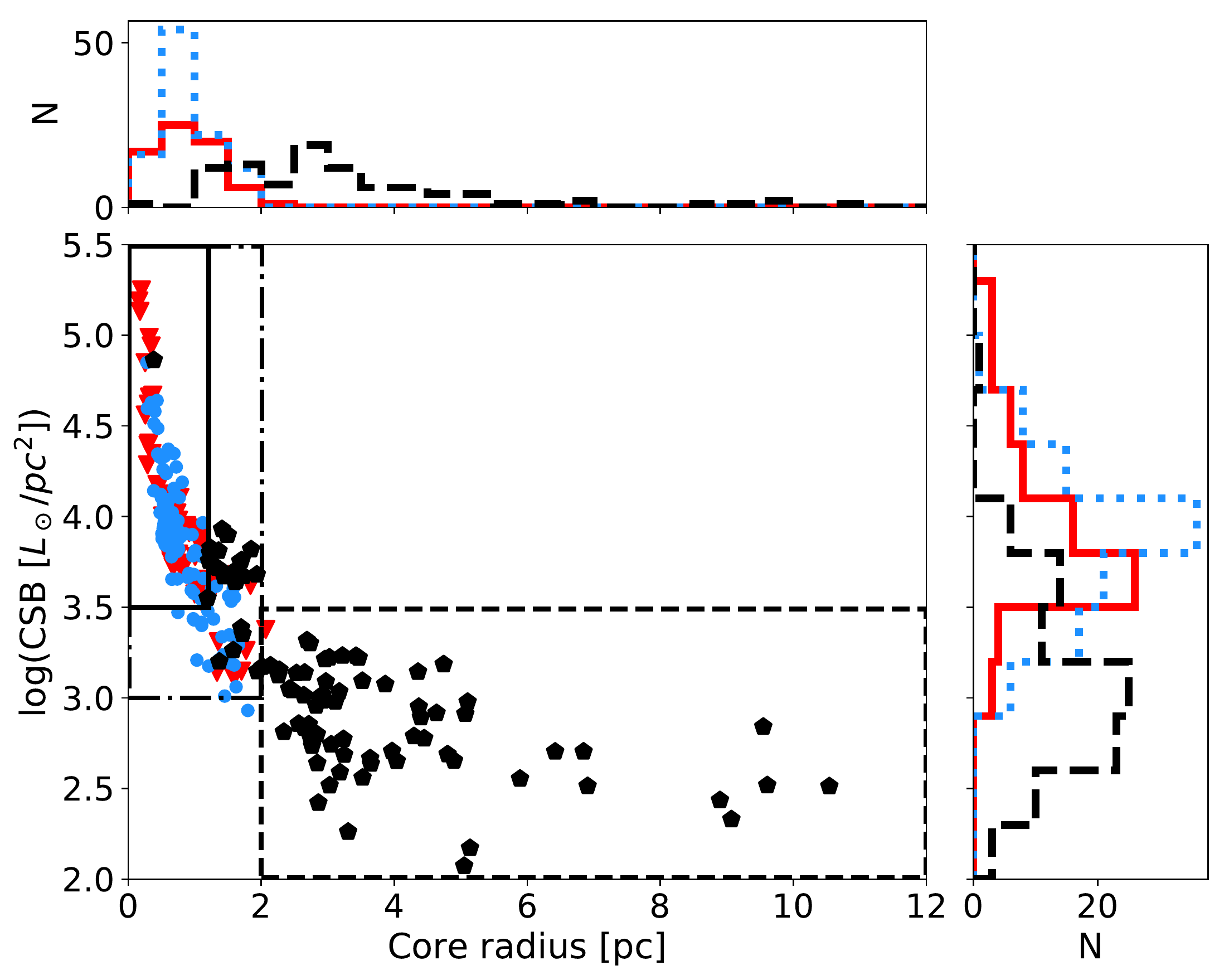}\\
  \end{subfigure}
  \begin{subfigure}{\columnwidth}
        \centering
        \includegraphics[width=\textwidth]{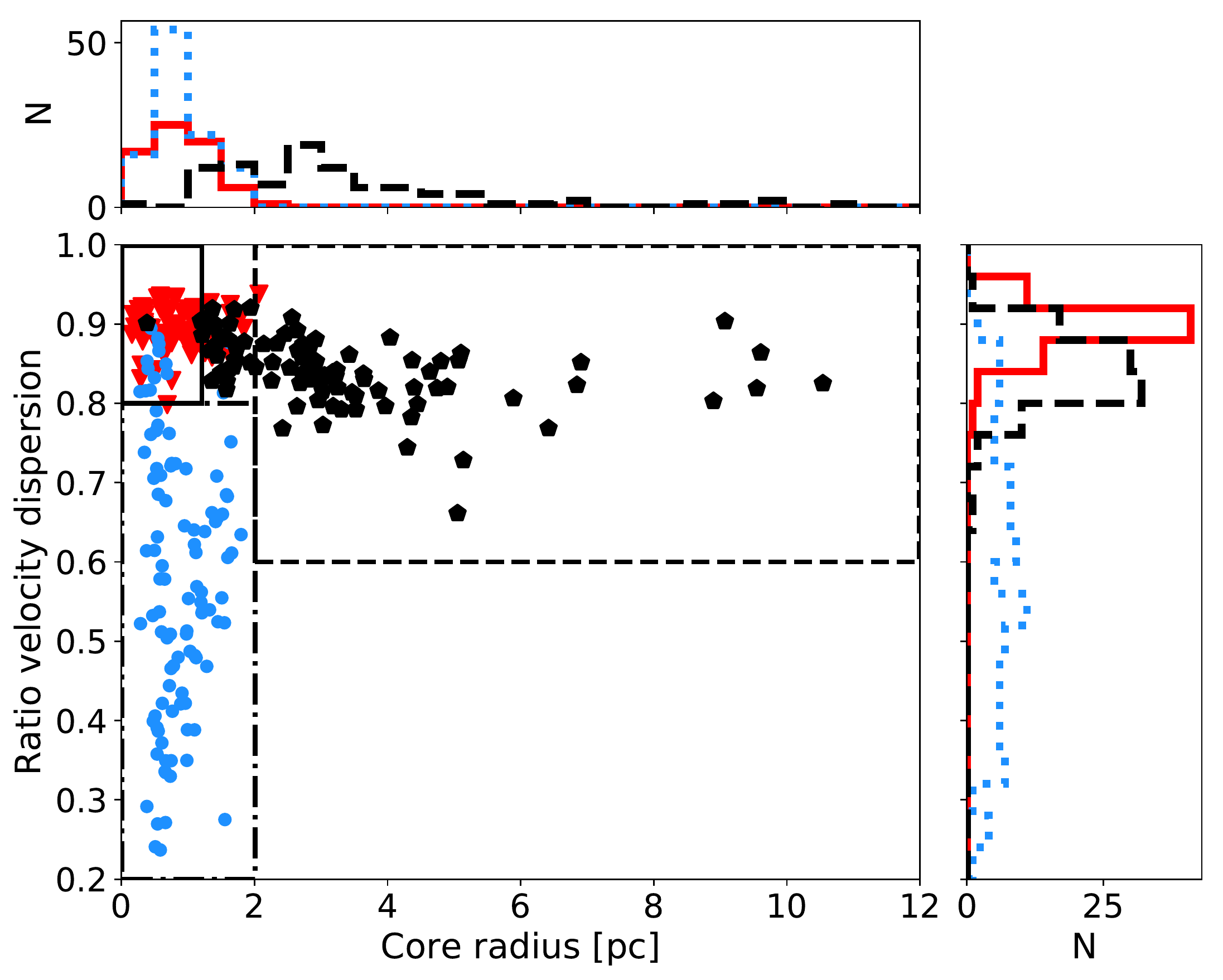}\\
        \hfill
        \centering
        \includegraphics[width=\textwidth]{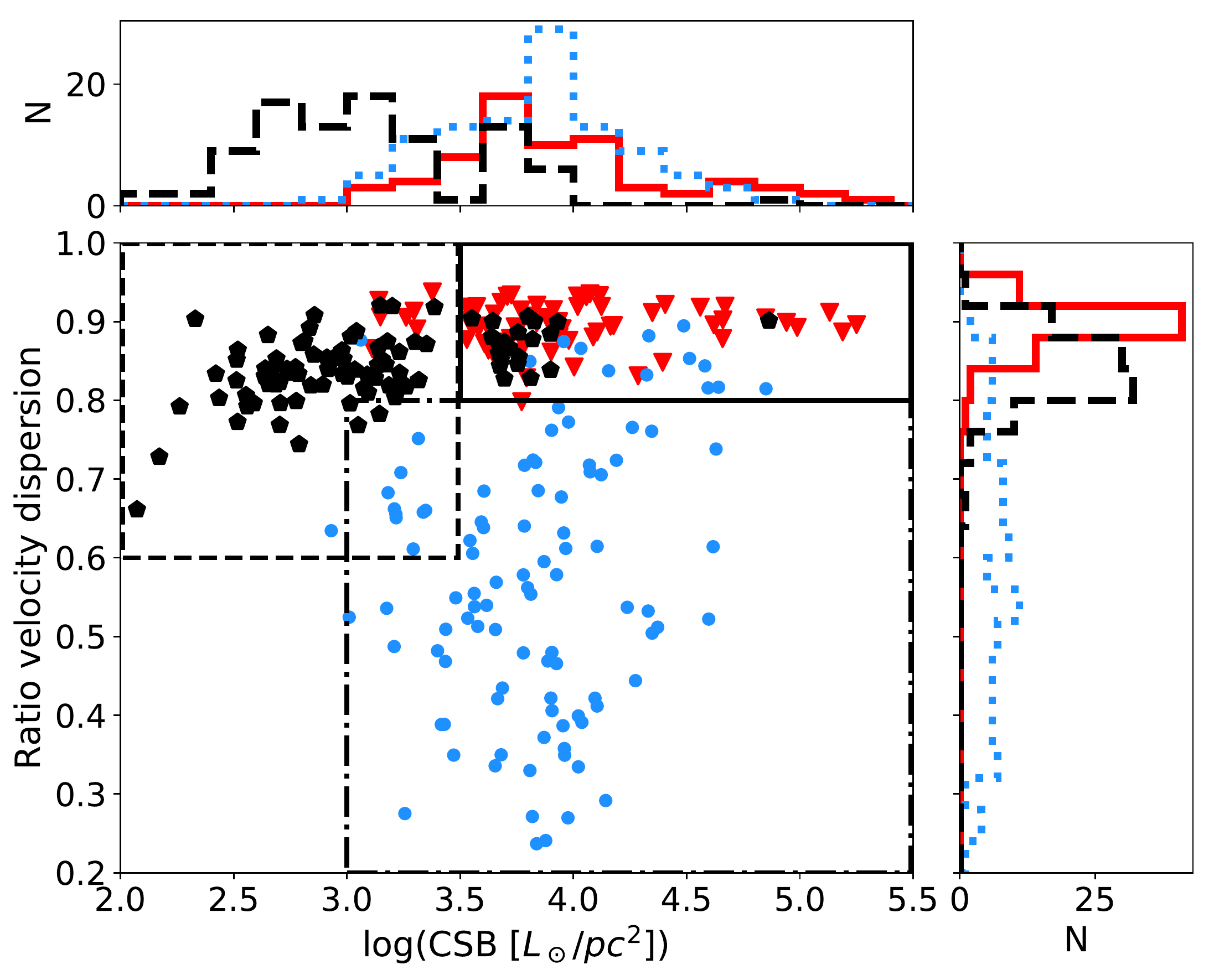}\\
  \end{subfigure}
  \caption{3D plot of our selected models, showing the $R_c$, the RVD and the CSB, for three dynamical models. On the side, the projection on three planes, with related histograms (red solid for Standard, blue dashed for IMBH and black dotted for BHS, respectively). Using three parameters it is possible to distinguish among different dynamical models. The continue line represents the border for the Region I, the dashed-dotted line for Region II, and the dotted line Region III respectively; see text for more details.}
    \label{Fig:3Dplot-csb-core-ratio}
\end{figure*}

Considering $R_c$, the ratio of radial velocity dispersion at $R_{hl}$ and the CRVD (RVD), and CSB, one can see that the three dynamical models are grouped in different regions of the space. This is clearly visible in Fig. \ref{Fig:3Dplot-csb-core-ratio}. The BHS models are mostly concentrated in large $R_c$ ($\gtrsim 2.0 \,pc$), relatively small CSB ($\lesssim 3.0 \,L_\odot/pc^2$) with RVD $\lesssim 0.9$; IMBH models are mostly concentrated in small $R_c$ ($\lesssim 2.0\, pc$), CSB $\gtrsim 3.5\,L_\odot/pc^2$ and small RVD ($\lesssim 0.9$); Standard models, instead, have small value of  $R_c$ ($\lesssim 2.0\, pc$), CSB $\gtrsim 3.0\,L_\odot/pc^2$ and  high RVD ($\sim 0.9$).

\begin{figure*}
\begin{subfigure}{\columnwidth}
        \centering
        \includegraphics[width=\textwidth]{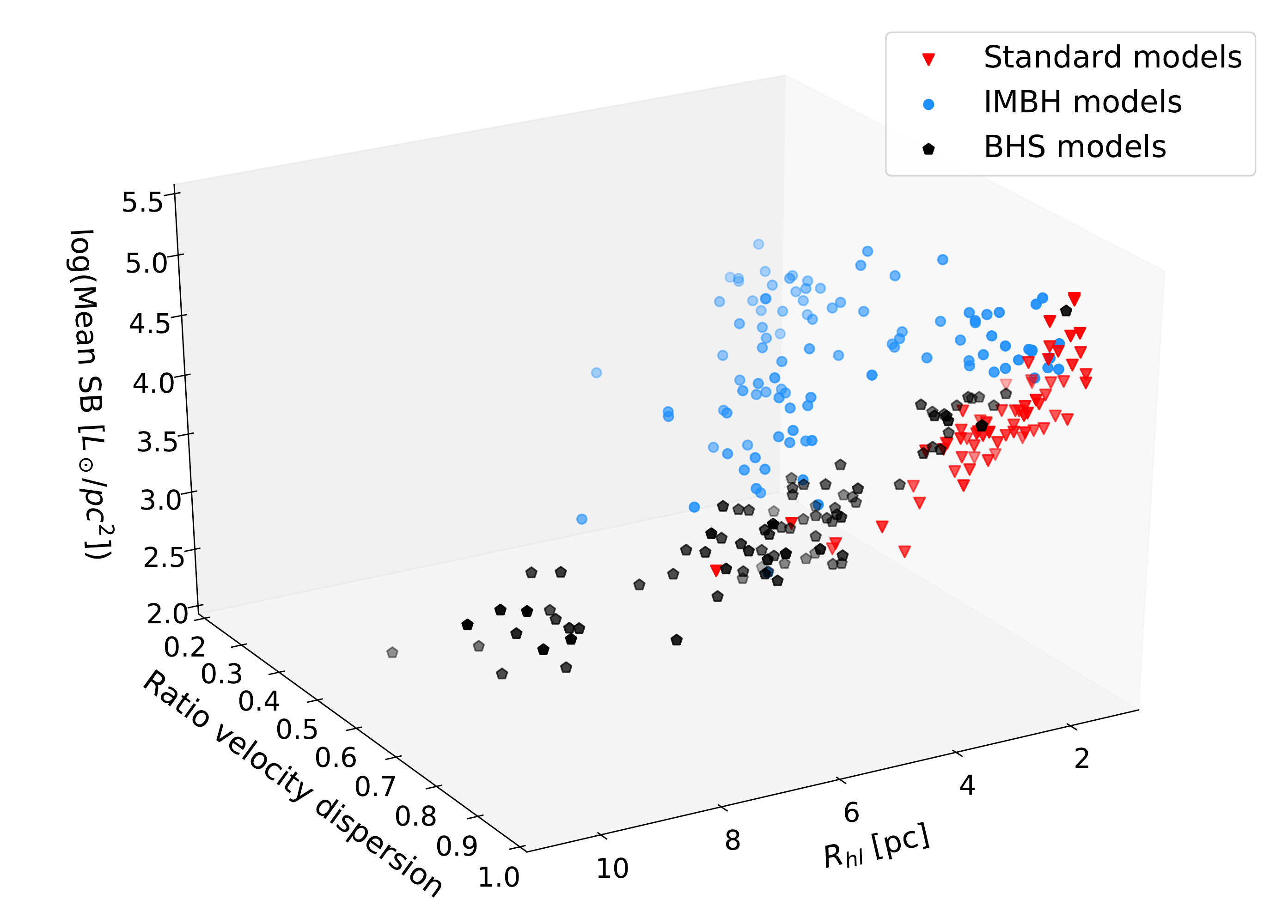}\\
        \hfill
        \vspace{10.5pt}
        \centering
        \includegraphics[width=\textwidth]{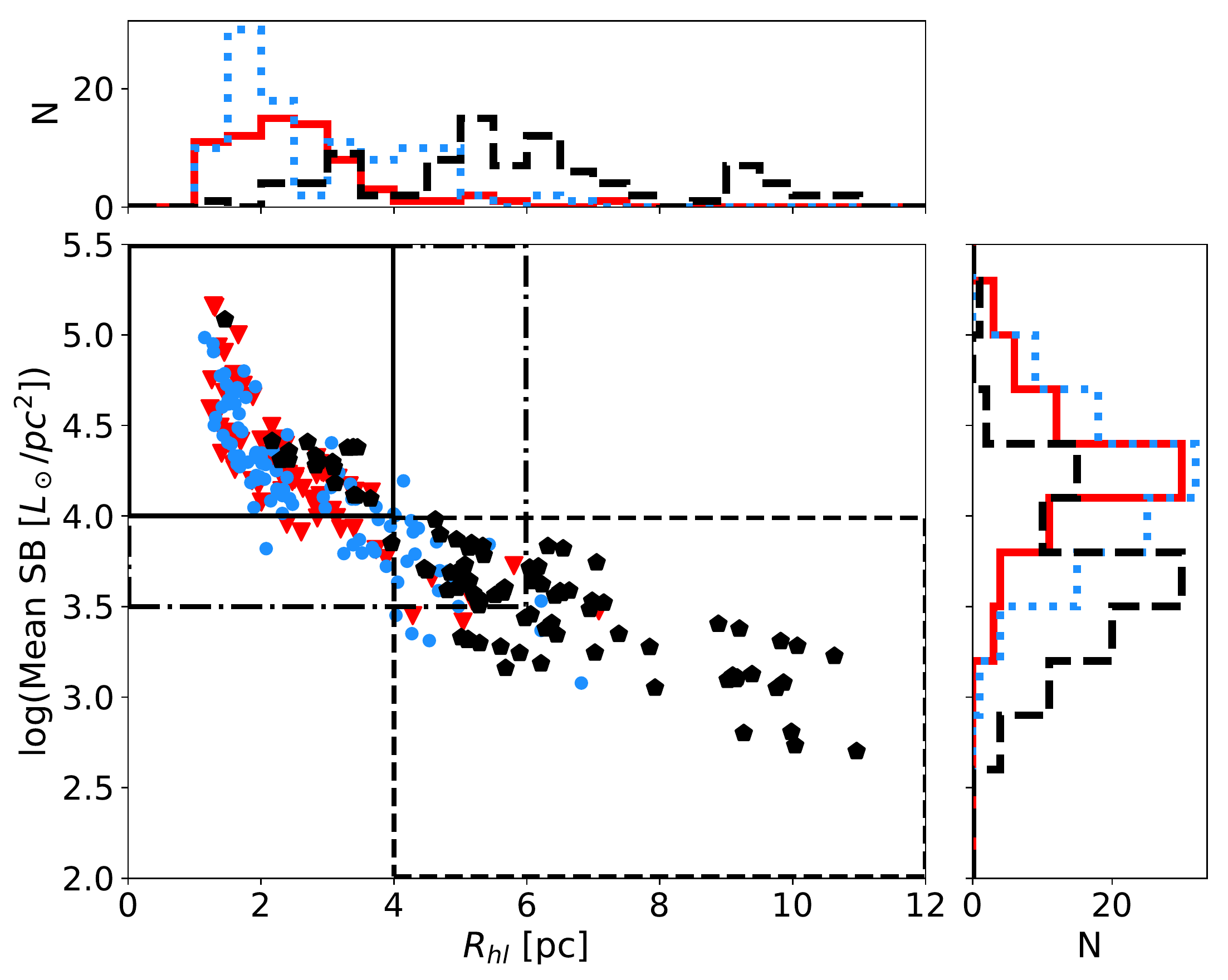}\\
  \end{subfigure}
  \begin{subfigure}{\columnwidth}
        \centering
        \includegraphics[width=\textwidth]{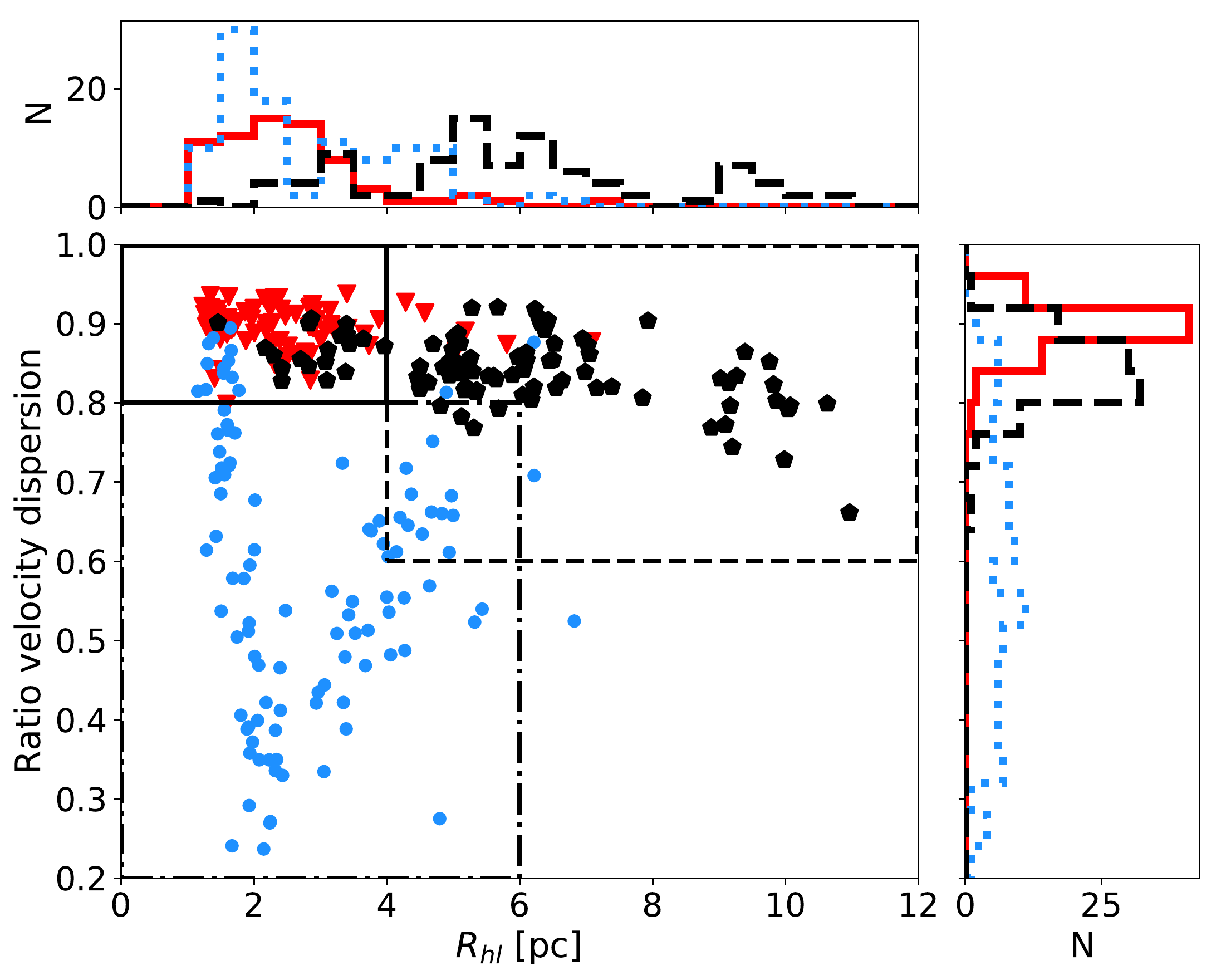}\\
        \hfill
        \centering
        \includegraphics[width=\textwidth]{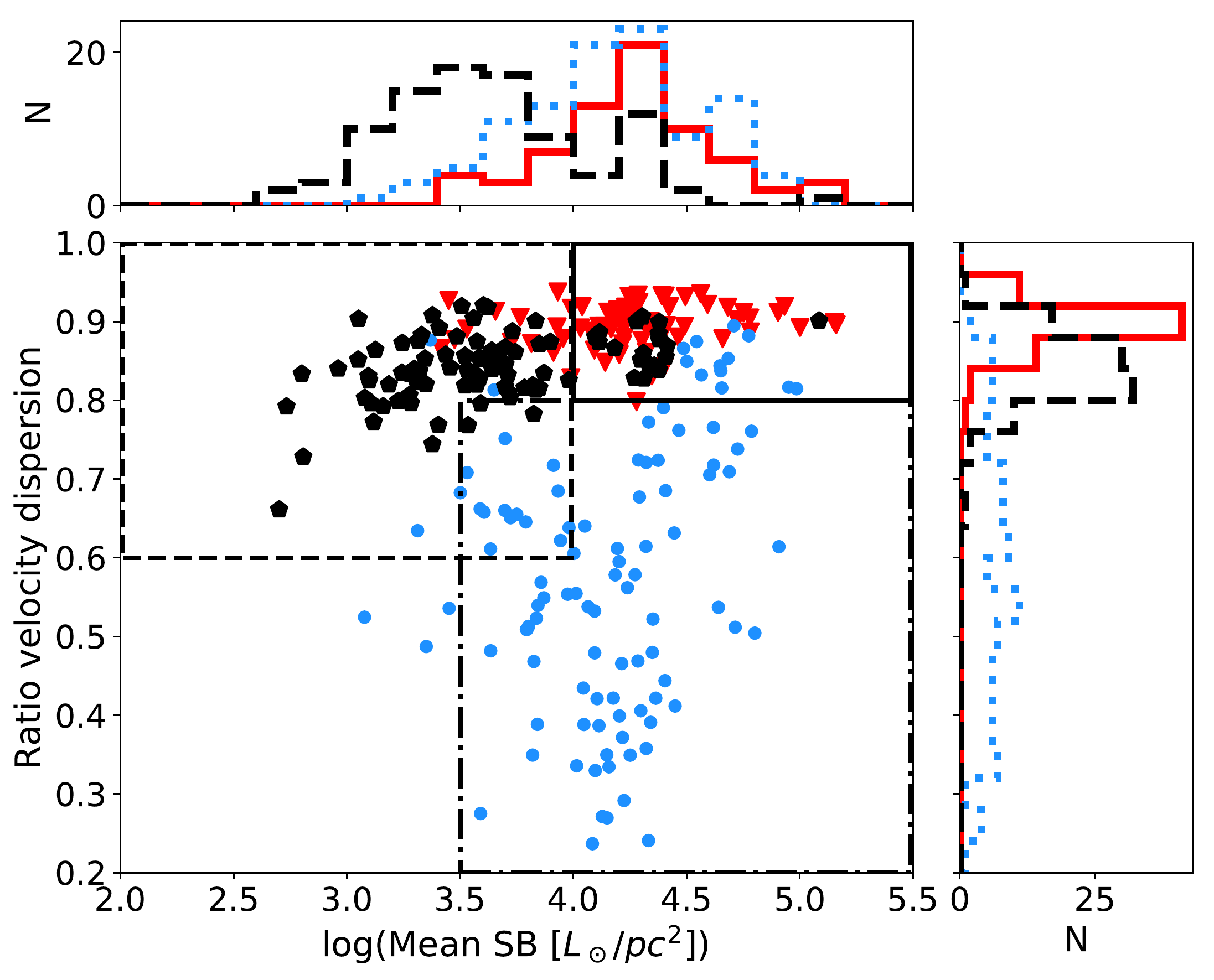}\\
  \end{subfigure}
  \caption{Same as in \ref{Fig:3Dplot-csb-core-ratio}, but the $R_{hl}$, the mean surface brightness and the RVD has been showed. }
    \label{Fig:3Dplot-meancsb}
\end{figure*}

\begin{figure*}
\begin{subfigure}{\columnwidth}
        \centering
        \includegraphics[width=\textwidth]{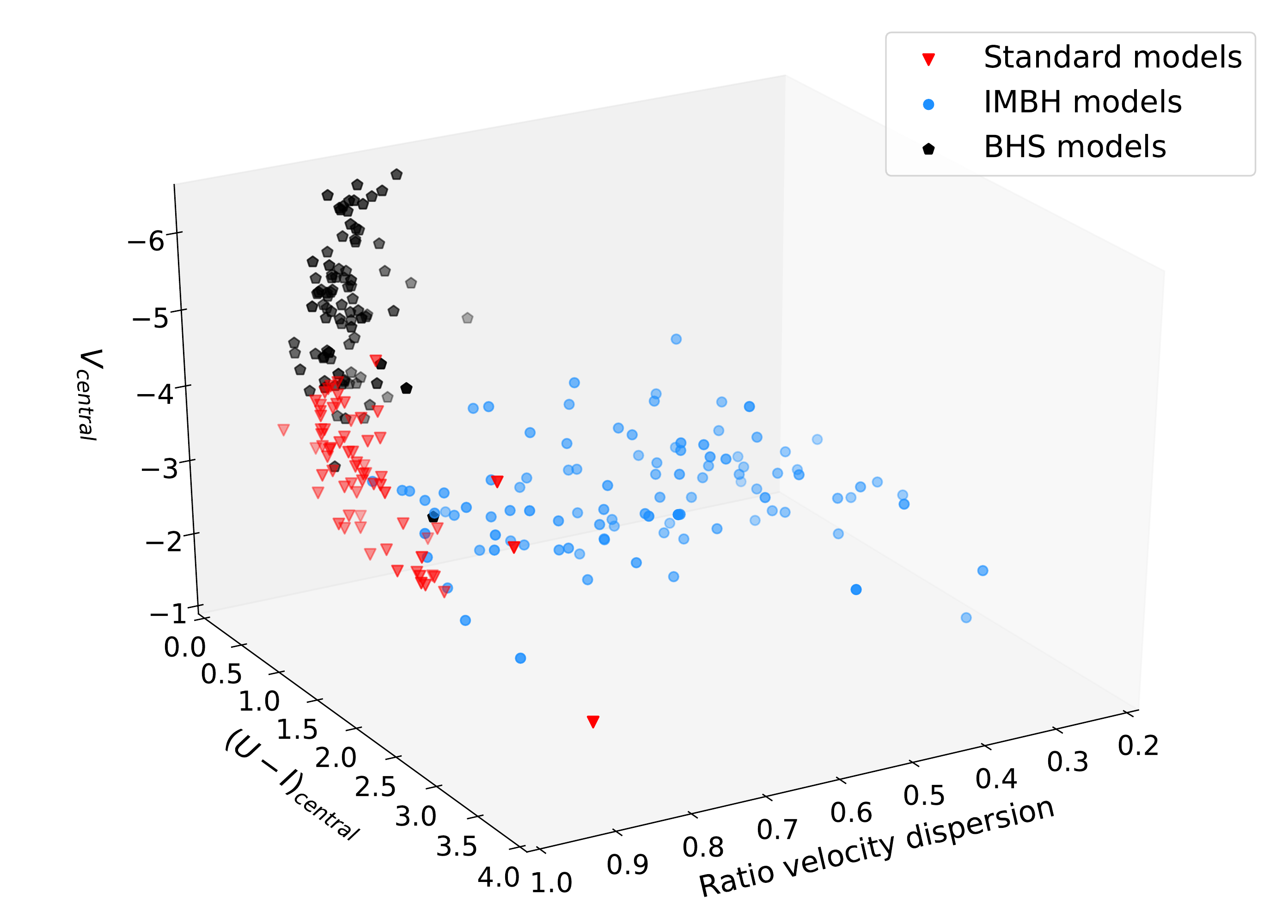}\\
        \hfill
        \vspace{10.5pt}
        \centering
        \includegraphics[width=\textwidth]{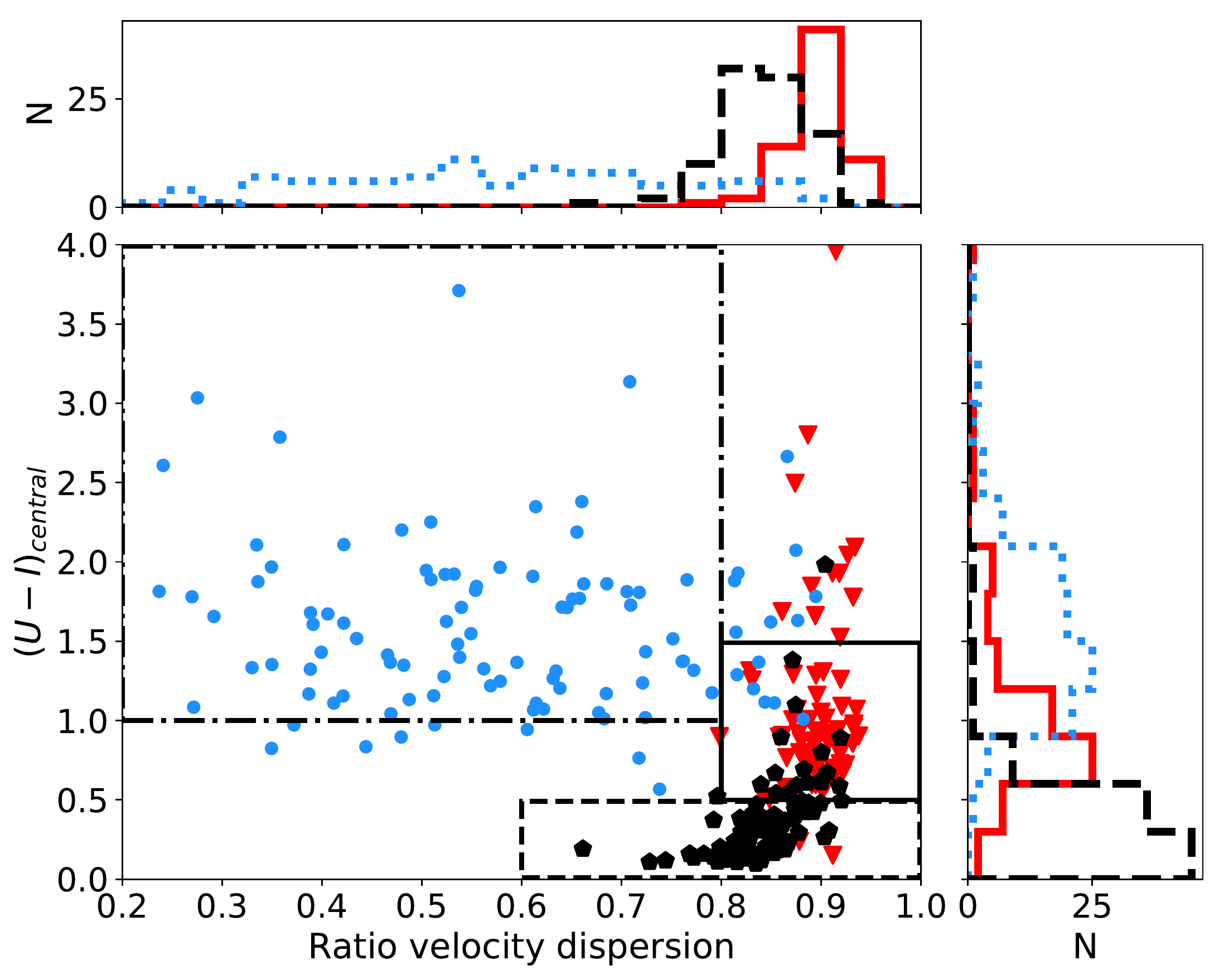}\\
  \end{subfigure}
  \begin{subfigure}{\columnwidth}
        \centering
        \includegraphics[width=\textwidth]{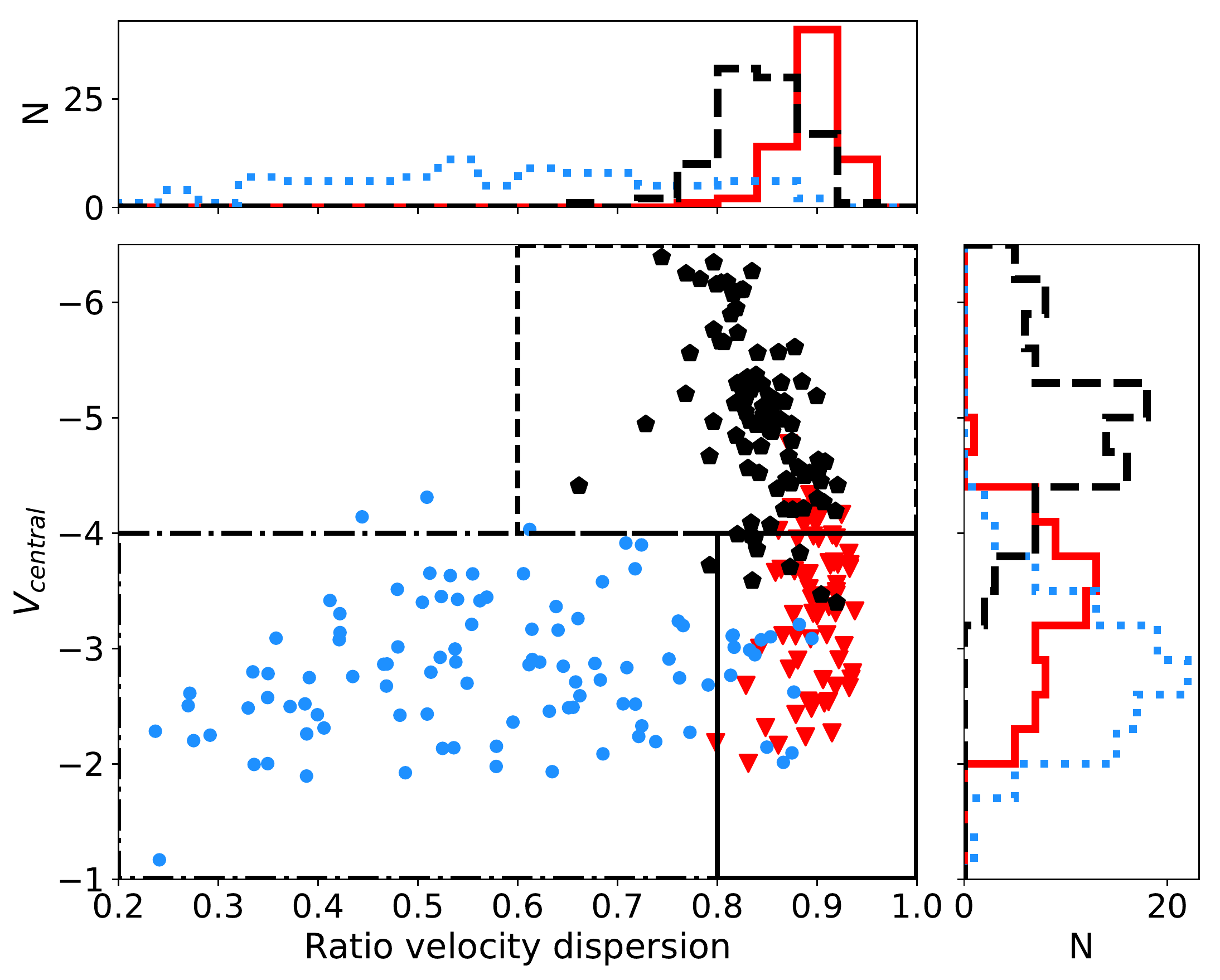}\\
        \hfill
        \centering
        \includegraphics[width=\textwidth]{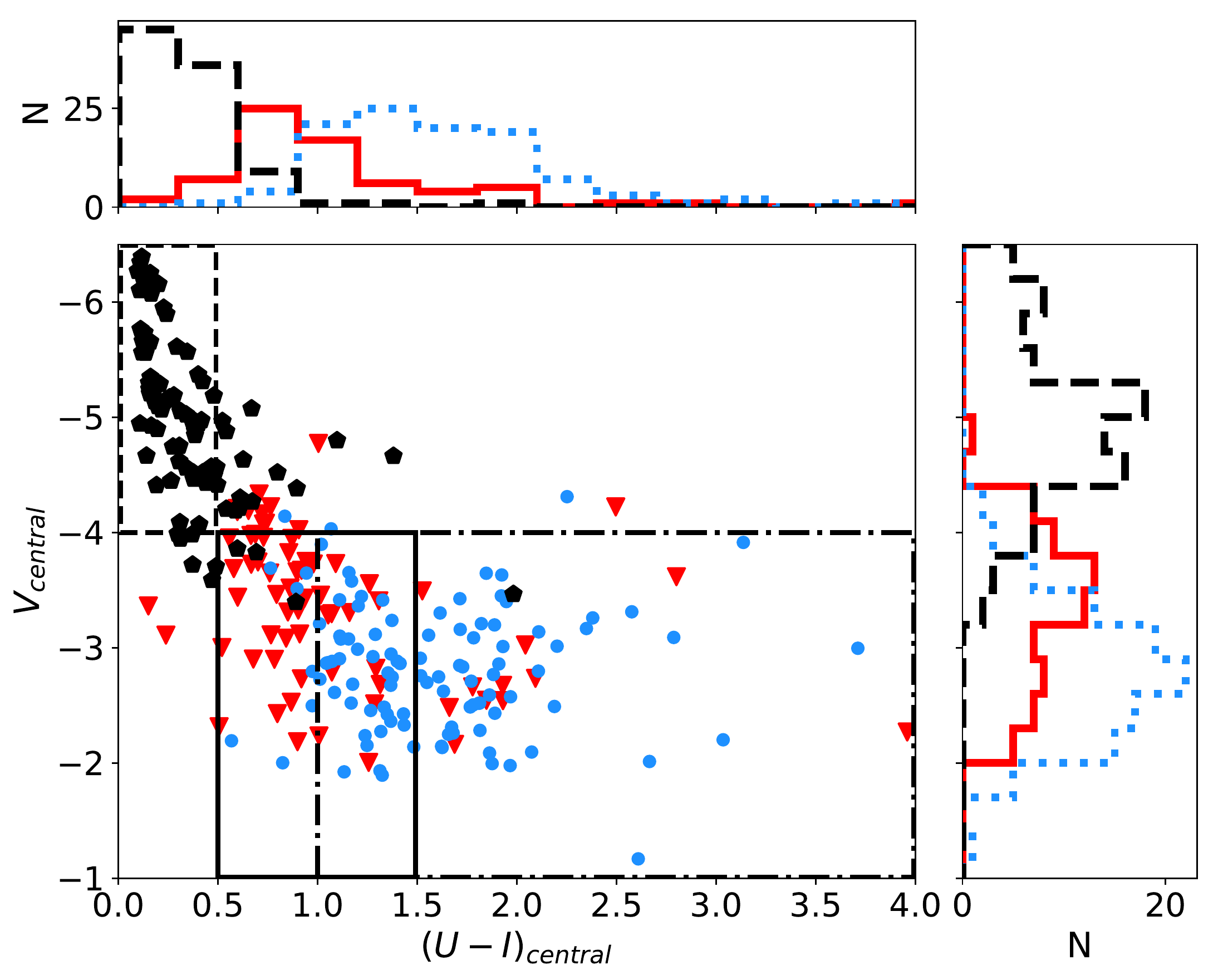}\\
  \end{subfigure}
  \caption{Same as in \ref{Fig:3Dplot-csb-core-ratio}, but the central $U-I$ color, the central $V$ magnitude and the RVD has been showed.}
    \label{Fig:3Dplot-UIV}
\end{figure*}

\begin{figure*}
\begin{subfigure}{\columnwidth}
        \centering
        \includegraphics[width=\textwidth]{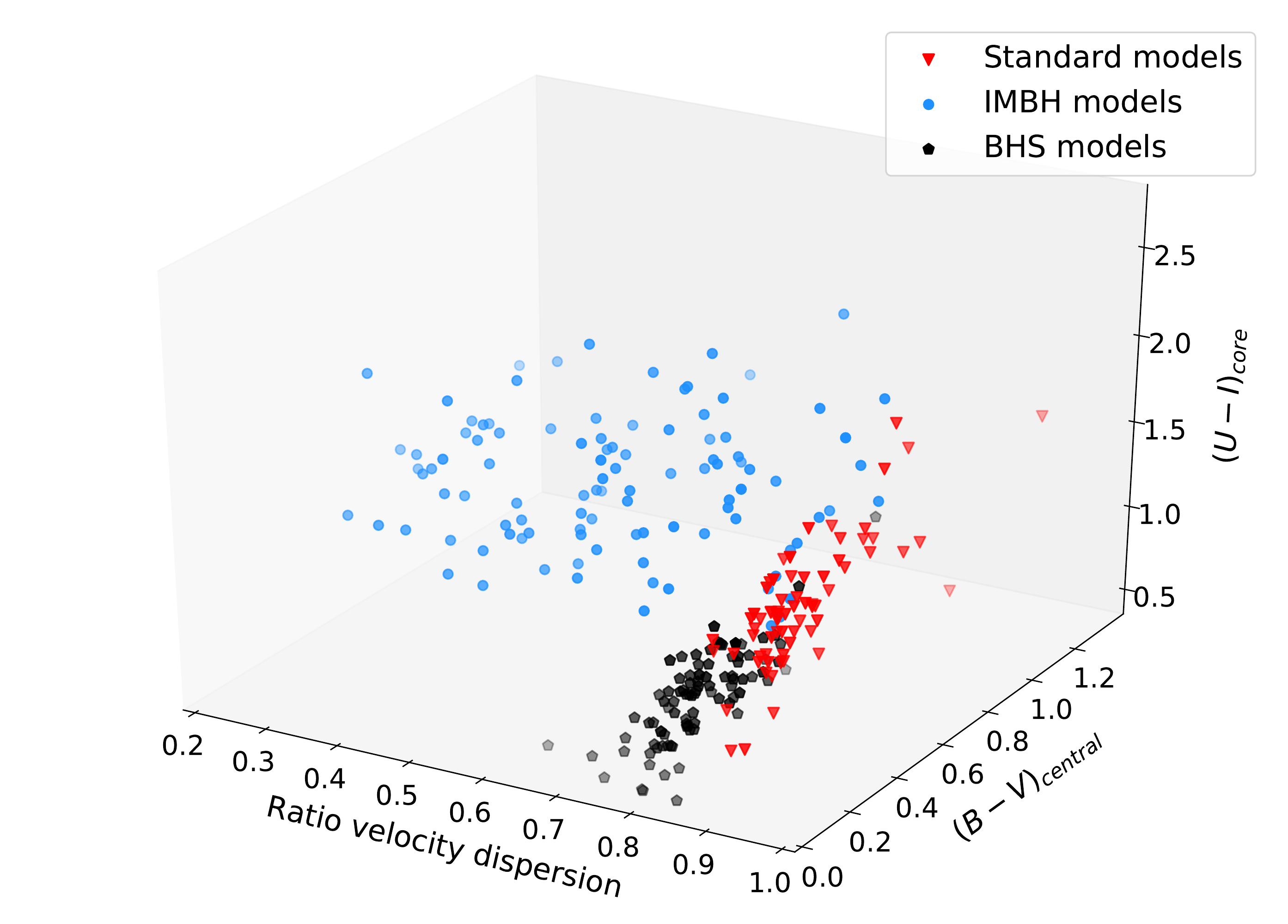}\\
        \hfill
        \vspace{10.5pt}
        \centering
        \includegraphics[width=\textwidth]{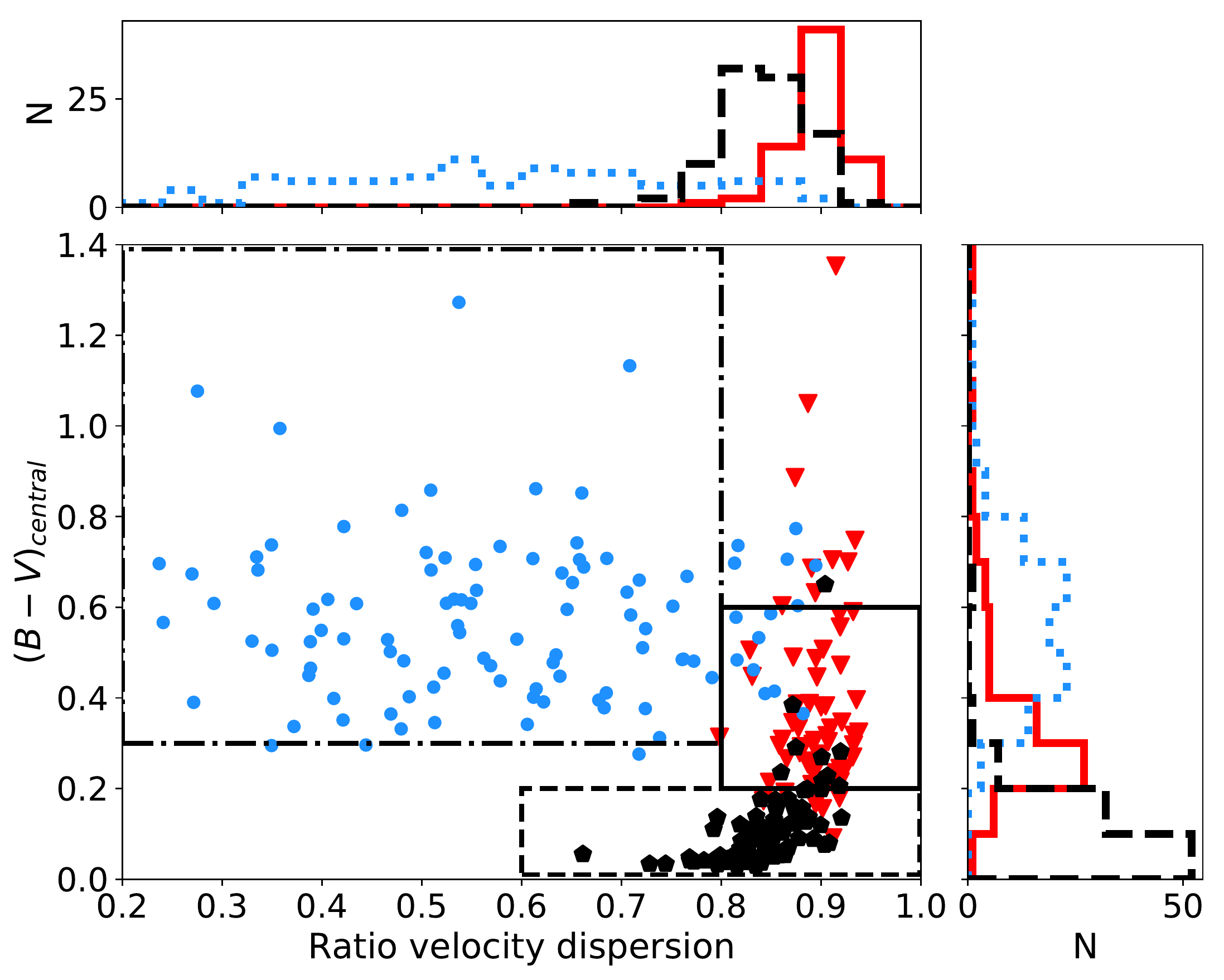}\\
  \end{subfigure}
  \begin{subfigure}{\columnwidth}
        \centering
        \includegraphics[width=\textwidth]{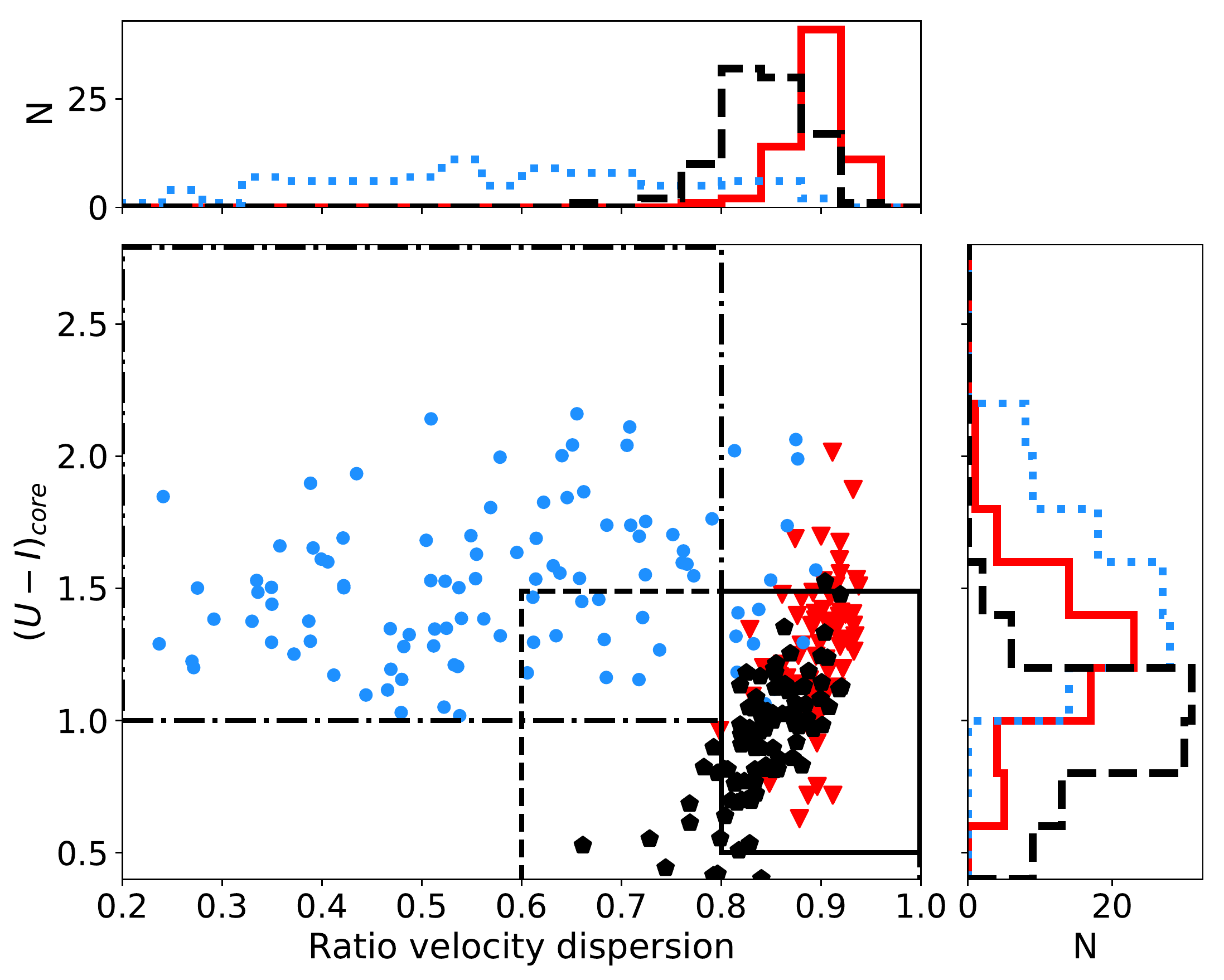}\\
        \hfill
        \centering
        \includegraphics[width=\textwidth]{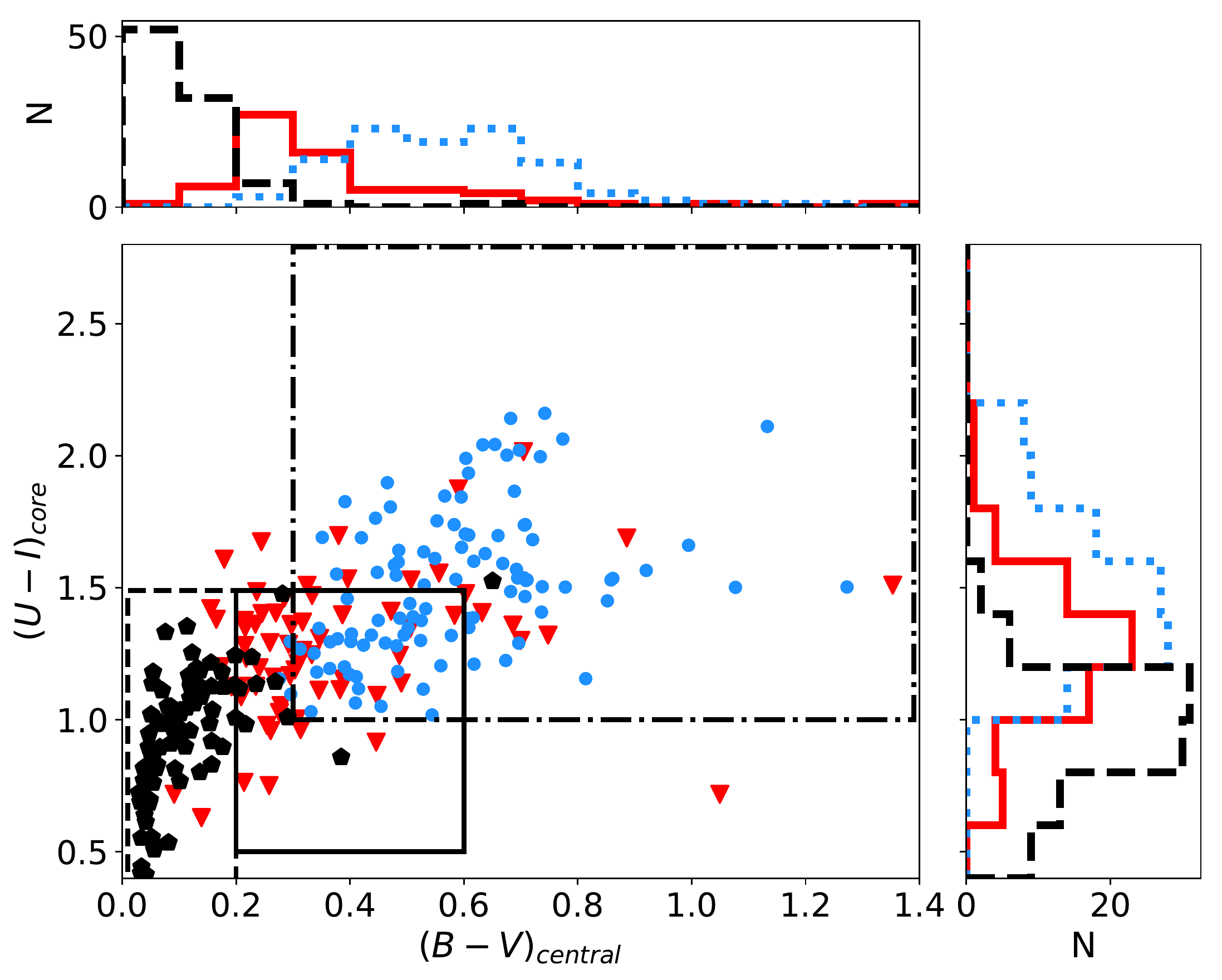}\\
  \end{subfigure}
  \caption{Same as in \ref{Fig:3Dplot-csb-core-ratio},  but the central $B-V$ color, the $U-I$ color at $R_c$, and the RVD have been showed.}
    \label{Fig:3Dplot-VBUI}
\end{figure*}

\begin{table*}
    \centering
    \begin{tabular}{ cccc}
    \hline
    Parameters & Region & Borders & Number of models\\
    \hline
    \multirow{9}{*}{$R_{c}, \, RVD, \, CSB$} &\multirow{3}{*}{I} &$R_c \leq 1.2 \,pc$ & \textbf{Standard: 57 (21.4)}\\
    & & $0.8 <$ RVD $  < 1.0$ & IMBH: 12 (4.5)\\
    & & $CSB \geq 10^{3.5} \,L_\odot/pc^2$ & BHS: 2 (0.75) \\
    \cline{2-4}
     & \multirow{3}{*}{II} & $R_c \leq 2.0\,pc$ &  Standard:  1 (0.4)\\
     & & RVD $ < 0.8$& \textbf{IMBH:  89 (33.4)} \\
     & & $CSB > 10^{3.0} \,L_\odot/pc^2$  &  BHS:  0 (0.0) \\
    \cline{2-4}
     & \multirow{3}{*}{III} &$R_c > 2.0\,pc$ & Standard:  1 (0.4)\\
     & & RVD $  < 1.0$ &  IMBH:  0 (0.0) \\
     & & $CSB < 10^{3.5} \,L_\odot/pc^2$ & \textbf{BHS:  67 (25.2)}\\
    \hline
     \hline
    \multirow{9}{*}{$R_{hl}, \, RVD, \, Mean\,SB$} &\multirow{3}{*}{I} &$R_{hl} < 4.0 \,pc$ & \textbf{Standard: 54 (20.3)}\\
    & & $0.8 <$ RVD $  < 1.0$ & IMBH: 12 (4.5) \\
    & & $Mean \, SB > 10^4 \,L_\odot/pc^2$ & BHS: 19 (7.1) \\
    \cline{2-4}
     & \multirow{3}{*}{II} &$R_{hl} < 6.0\,pc$ & Standard: 1 (0.4)\\
     & & RVD $ < 0.8$ & \textbf{IMBH: 85 (31.9)} \\
     & & $Mean\, SB \geq 10^{3.5} \,L_\odot/pc^2$ & BHS: 3 (1.1) \\
    \cline{2-4}
     & \multirow{3}{*}{III} &$R_{hl} >  4.0\,pc$ & Standard: 5 (1.9)\\
     & & RVD $  < 1.0$ & IMBH: 17 (6.4) \\
     & & $Mean\,SB <  10^{4} \,L_\odot/pc^2$ & \textbf{BHS: 73 (27.4)}\\
    \hline
     \hline
    \multirow{9}{*}{$(U-I)_{central}, \, RVD, \, V_{central}$} &\multirow{3}{*}{I} &$0.5 < (U-I)_{central} < 1.5$ & \textbf{Standard: 44 (16.5)}\\
    & & $0.8 <$ RVD $  < 1.0$ & IMBH: 6 (2.2) \\
    & & $V_{central} > -4.0$ & BHS: 3 (1.1)\\
    \cline{2-4}
     & \multirow{3}{*}{II} &$(U-I)_{central} > 1.0$ & Standard: 0 (0.0)\\
     & &  RVD $ < 0.8$ & \textbf{IMBH: 80 (30.1)} \\
     & & $V_{central}  > -4.0$ & BHS: 0 (0.0) \\
    \cline{2-4}
     & \multirow{3}{*}{III} &$(U-I)_{central} < 0.5$ & Standard: 0 (0.0)\\
     & &  RVD $  < 1.0$ & IMBH : 0 (0.0)\\
     & & $V_{central} < -4.0 $ & \textbf{BHS: 69 (25.9)}\\
    \hline
     \hline
    \multirow{9}{*}{$(B-V)_{central}, \, RVD, \, (U-I)_{core}$} &\multirow{3}{*}{I} &$0.2< (B-V)_{central} < 0.6$ & \textbf{Standard: 45 (16.9)}\\
    & & $0.8 <$ RVD $  < 1.0$ & IMBH: 7 (2.6) \\
    & & $0.5 < (U-I)_{core} < 1.5$ & BHS: 8 (3.0)\\
    \cline{2-4}
     & \multirow{3}{*}{II} &$(B-V)_{central} > 0.3$ & Standard: 0 (0.0)\\
     & &  RVD $ < 0.8$ & \textbf{IMBH: 87 (32.7)} \\
     & & $ (U-I)_{core} > 1.0$ & BHS: 0 (0.0) \\
    \cline{2-4}
     & \multirow{3}{*}{III} &$(B-V)_{central} < 0.2$ & Standard: 4 (1.5)\\
     & & RVD $  < 1.0$ & IMBH: 0 (0.0) \\
     & & $ (U-I)_{core} < 1.5$ & \textbf{BHS: 81 (30.4)}\\
    \hline
    
    \end{tabular}
    \caption{The definitions for the three different regions for each 3D space defined in this work. The first column names the three parameters that have been used; the second column shows the regions in which it has been divided; the third column indicates the border of each region. In the forth column, the number of dynamical models in each region are reported, respectively; in parenthesis the percentage of the model in the particular region compared to the total number of considered models (266) is reported too. In bold the model type which dominate in each region.}
    \label{Table:Limits}
\end{table*}

In Fig. \ref{Fig:3Dplot-meancsb} we show the same 3D plot, but we considered instead the $R_{hl}$, the mean surface brightness, and the RVD. The mean surface brightness is defined as the total luminosity of the system, divided by the area inside $R_{hl}$. Comparing Fig. \ref{Fig:3Dplot-meancsb} with Fig. \ref{Fig:3Dplot-csb-core-ratio}, we can see that the properties at $R_{hl}$ are not as good as the central properties for distinguish among the dynamical models, but still good enough to roughly separate the models. 

In Fig \ref{Fig:3Dplot-UIV}, we considered the central $U-I$ color, the central $V$ magnitude, and the RVD. It is possible to note that the central color and central magnitude can give a good distinction among dynamical models just as well. The same is true if we consider the central $B-V$ color, the $U-I$ color at $R_c$  and the RVD, as one can see in Fig. \ref{Fig:3Dplot-VBUI}.

For each of these 3D plots, we can divide the space in three regions, that would include as many models as possible of only one dynamical state, with the less contamination from others. The borders and the number of models in each region are listed in Table \ref{Table:Limits}. The division between the different regions has been chosen by eye. In \Cref{Fig:3Dplot-csb-core-ratio,Fig:3Dplot-meancsb,Fig:3Dplot-UIV,Fig:3Dplot-VBUI} we showed the boundaries for the three different regions. When considering the 2D projections for each figure, it is not possible to properly separate the three dynamical models in three distinct regions without the presence of dynamical contaminants for any of the figures. However, in the 3D plots such contaminants are relatively small and they will not strongly affect the statistical distinctions between different GC evolutionary models.

The best combination of properties seem to be $R_c$, CSB and RVD, or $(B-V)_{central}$, $(U-I)_{core}$ and RVD. This could be useful from an observational point of view: when the structural parameters of the system (such as $R_c$  or CSB) are not possible to determine, it is still possible to distinguish the different dynamical models using the central colors and magnitudes (if they can be observed). However, it is generally easier to determine and to observe properties at $R_{hl}$; even if the properties at this radius are not as good as the central values, they are still good enough to differentiate the dynamical models, and so they could be easier to verify.

\section{Discussion} \label{sec:Discussion}

\cite{Askar2017} already showed that GCs models from the MOCCA-Survey Database results relatively well represent MWGCs from  \citet[updated 2010]{Harris1996}. In the first part of this paper, we showed that our selected subsample is also able to reproduce observed properties of MWGCs. In particular, in Sec \ref{sub-sec:MWGCs}, we also compared our database with the \cite{Baumgardt2018} catalog. The overall result is that our subsample reproduces the Galactic GCs' global properties' distributions. Small differences have been noticed: our sample shows a higher mean $R_c$  when compared to both catalogs and our mean CRVD is smaller than that of the Baumgardt catalog's. Those differences can be due to observational errors and some systematic shifts connected with different techniques used to analyze the observational and simulation data. However, this gave us the confidence that our method and subsample could be applied also to EGGCs, with the assumption that all old GCs were formed in similar physical environments and similar physical processes were responsible for their formation.

In Sec. \ref{sub-sec:comp-prev-works} we compared the results for Galactic GCs from our subsample with the results of the considered previous works \citep{Askar2018,Weatherford2019,ArcaSedda2019}. Firstly, it would be important to list the differences in models selection used by the different groups. \cite{Askar2018} and \cite{ArcaSedda2019} also used the MOCCA-Survey Database results to identify BHS and IMBH models respectively, but with a different selection criteria. The authors in  \cite{Askar2018} selected models accordingly to their CSB and the observed present-day half-mass relaxation time. In \cite{ArcaSedda2019}, the authors labelled a Galactic GCs as IMBH (or BHS) according to how many MOCCA models (among the 10 closest in 6D observational parameter space) contain an IMBH (or a BHS respcetively). For this purpose, they used as properties the visual and bolometric total luminosity, half-mass and core radii,  galactocentric  distance,  average  and  central  surface luminosity. We would like to underline that in both those two works, the authors used all of the models from the MOCCA-Survey Database, also those models with no mass fallback prescription and tidally filling models (which we exclued, as explained in Sec. \ref{subsec:model-selection}). Instead, in \cite{Weatherford2019}  the authors correlates the number of BHs in the system with the mass segregation parameter $\Delta$, obtained from the 2D-projected snapshots of models presented in the CMC Cluster Catalog \citep{Kremer2020}. They did not use any other parameters to distinguish between IMBH, Standard and BHS models. As was shown in Section \ref{subsec:3D}, at least only three observational parameters can guarantee a relatively solid distinction between different cluster evolution histories. 

A difference in the number of Galactic GCs that have been classified as IMBH or BHS in previous works and in this one can be partially explained by too strong and conservative of a limitation we have used in this paper (see definition of ``small'' and ``large core radii'' clusters, respectively, in Sec. \ref{sub-sec:comp-prev-works}): we have considered only observed GCs with $M_V <-6.5$, for which all properties are available, and a conservative choice of a minimum $R_c$  of 2.5 pc. The reason for this choice is to consider two regions of the GCs' properties ($R_c$, CRVD, CSB) that would mostly contain the two dynamical models in interest (IMBH and BHS), reducing as much as possible the region where those two could overlap with Standard models' properties. Moreover, we want to use parameters and regions that would be relatively easy to define (and to observe) for EGGCs, that are the main target of our project. Additionally, in \cite{Askar2018} (see also \cite{ArcaSedda2018}), the  limitations are less conservative. The authors considered all clusters with  CSB $ < 10^4\,\, L_\odot/pc^2$ and observed present-day half-mass relaxation times $< 0.9 \,\,Gyr$. This is clearly visible in Fig. \ref{Fig:comparison} (on the left side): if we were to consider clusters with radius $\ge 2.0\,\,pc$, irregardless of whether CRVD information is available or not, the number of ``large core radii'' models would increase up to $\sim 17$.

We also compared our ``small core radii'' clusters (that correspond mostly to IMBH), with the list of GCs reported in \cite{ArcaSedda2019}. In their work, 35 observed Galactic GCs are likely to harbour an IMBH, meanwhile in our work only 15 do. In this case, the main difference is our observational limit we imposed on our models ($M_V <-6.5$): by considering only those that satisfied this limit, their number of shortlisted GCs decreases to 17. Again, if we would consider clusters independently of all information being available, the number of GCs that we shortlisted as possibly harbouring an IMBH will increase up to $\sim 30$. In a photometric and spectroscopic study, \citep{Lutzgendorf2013} listed a sample of 13 IMBH MWGC candidates, a similar number reported in this paper. In \ref{Table:strongly-concentrated} we also report the cluster listed in that work.

The definition of ``small'' and ``large core radii'' clusters could be useful, from an observational point of view, for the identification and classification of the dynamical state of real GCs, knowing only the global properties of the system. 

Finally, we showed that different types of observational cluster parameters connected with the central properties are needed to distinguish with more confidence between different cluster dynamical evolutions. In particular, as showed in Sec. \ref{subsec:3D}, a minimum of three parameters is necessary (two parameters are not enough, see Fig. \ref{Fig:color-gradient}). This could be explained by the nature and structure of the systems due to their dynamical histories.

Indeed, it is expected that the influence of the IMBH would change the central properties of the GC: due to the deeper central potential, the system is expected to be more concentrated (small $R_c$), having a high CRVD (RVD $<0.8$) and a high CSB $CSB > 10^3 \,L_\odot/pc^2$.

The central part of a BHS model is dominated by the presence of BHs. This implies that the observed $R_c$  for such a system is expected to be larger (it is measured considering only luminous stars). The presence of a BHS in the central part of the system would imply a relatively small CSB, since the core of the system is dominated by the not luminous
BHs. However, a strong gradient in velocity dispersion is not expected (implying a RVD $\sim 0.9$), since we have considered BHS models with systems that are in balanced evolution: indeed, the BHS is not detached from the whole system \citep{BreenandHeggie2013b} and the $R_c$  is not strongly different from the half-mass radius.

The Standard models, finally, are models expected to be approaching core collapse or in the post-collapse phase. This means a small $R_c$ and relatively large $R_h/R_c$ ratio ($R_c < 2.0\,pc$, $CSB > 10^3 \,L_\odot/pc^2$) and  RVD is $\sim 0.9$ (or smaller).

\begin{figure*}
\begin{subfigure}{\columnwidth}
    \centering
    \includegraphics[width=\textwidth]{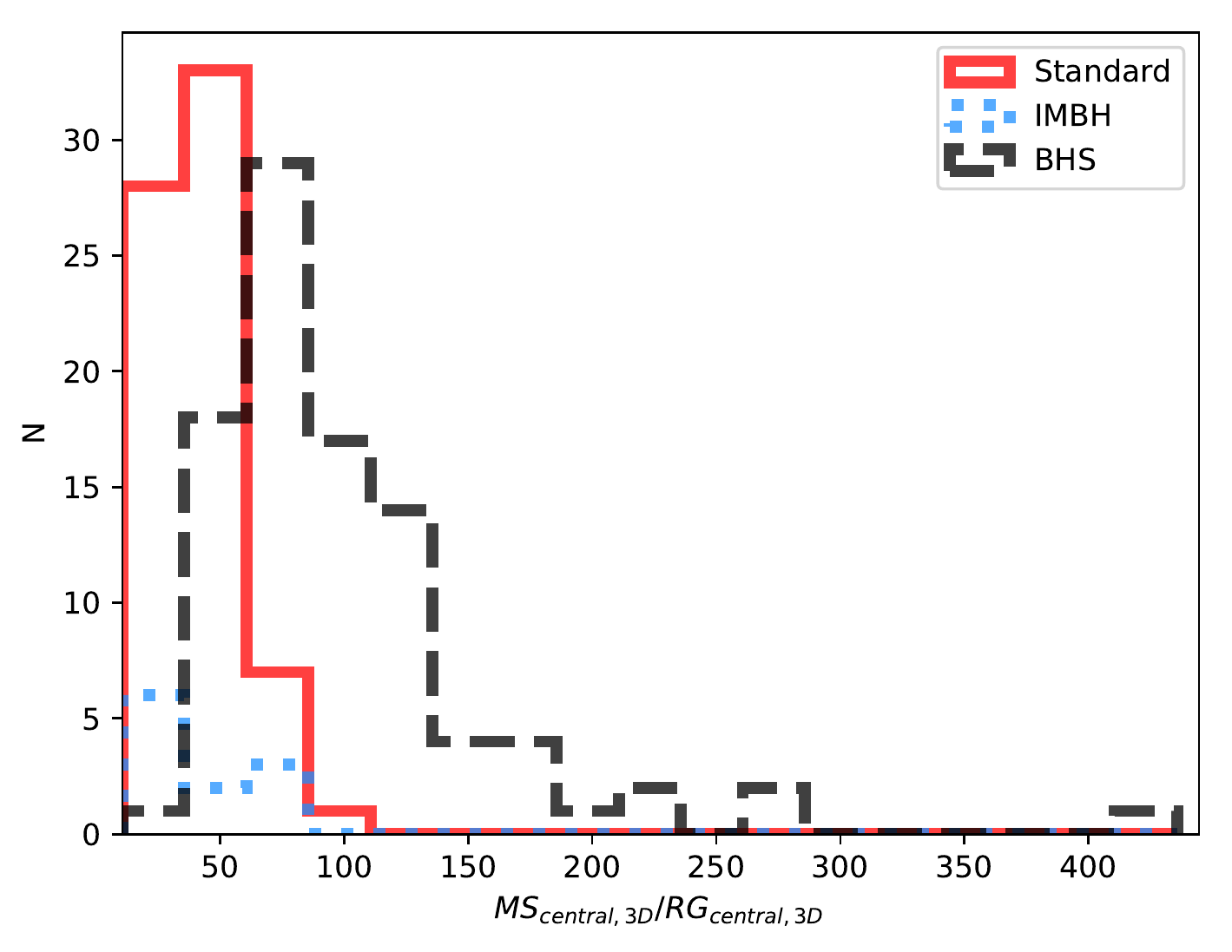}
    \hfill
    \includegraphics[width=\textwidth]{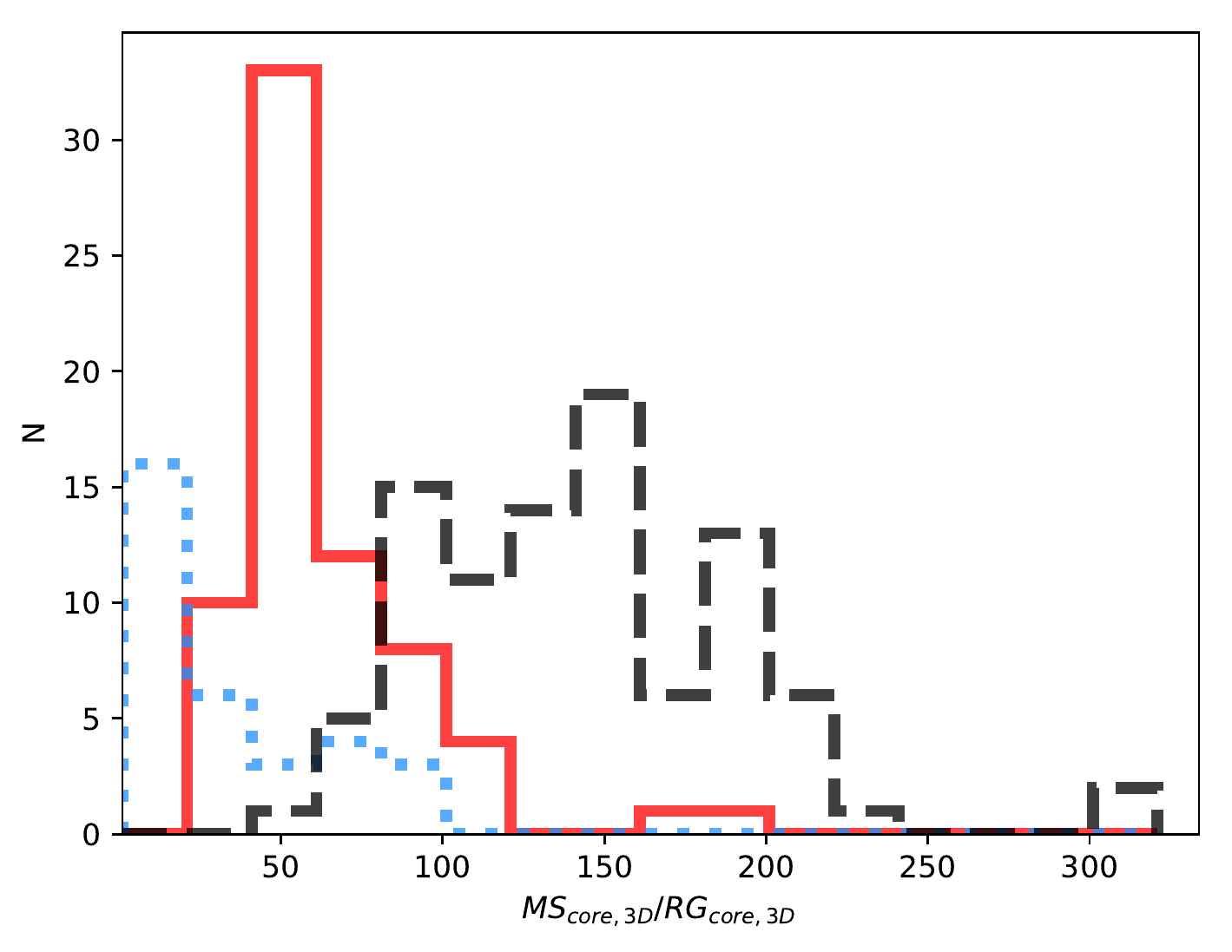}
  \end{subfigure}
  \begin{subfigure}{\columnwidth}
    \centering
    \includegraphics[width=\textwidth]{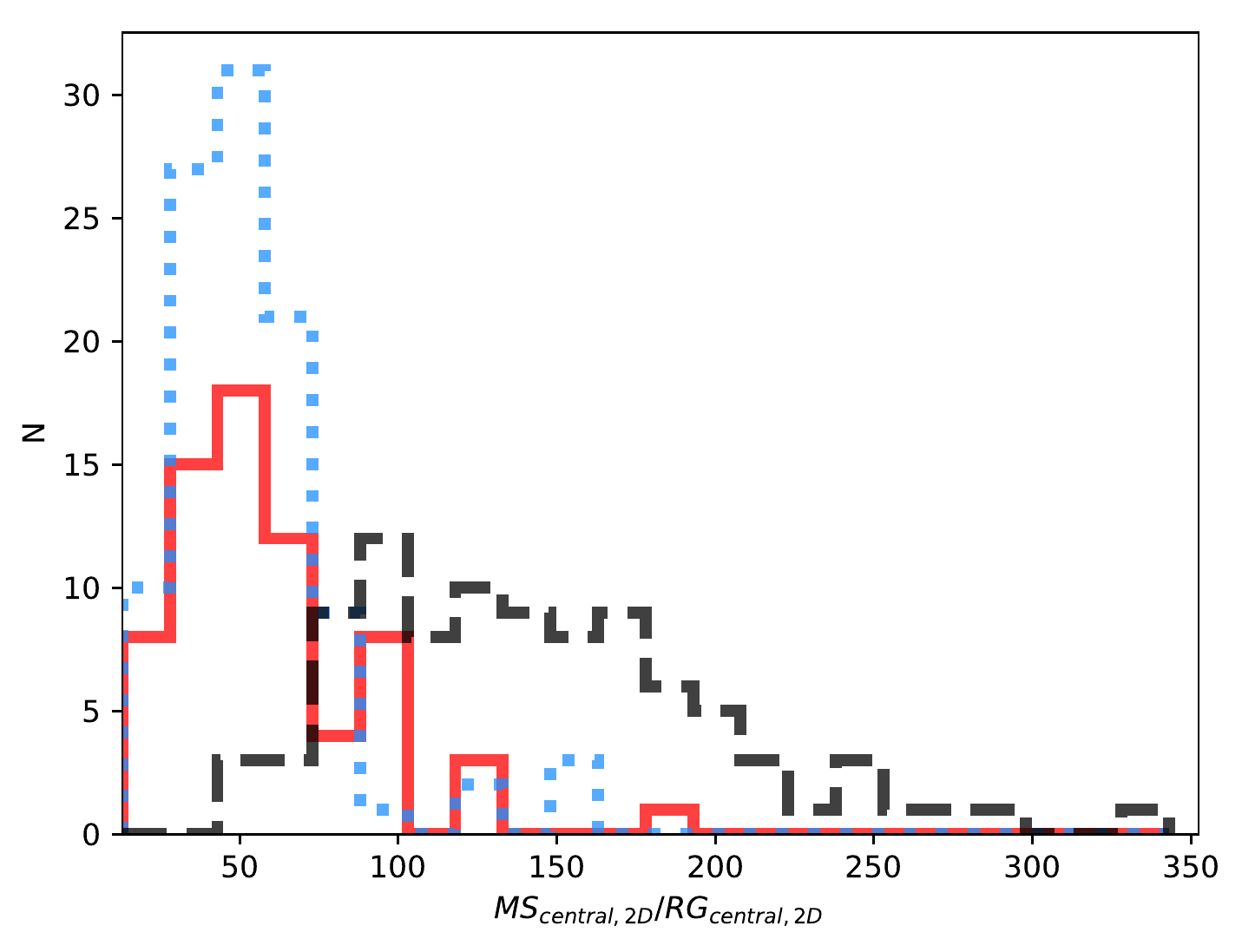}
    \hfill
    \includegraphics[width=\textwidth]{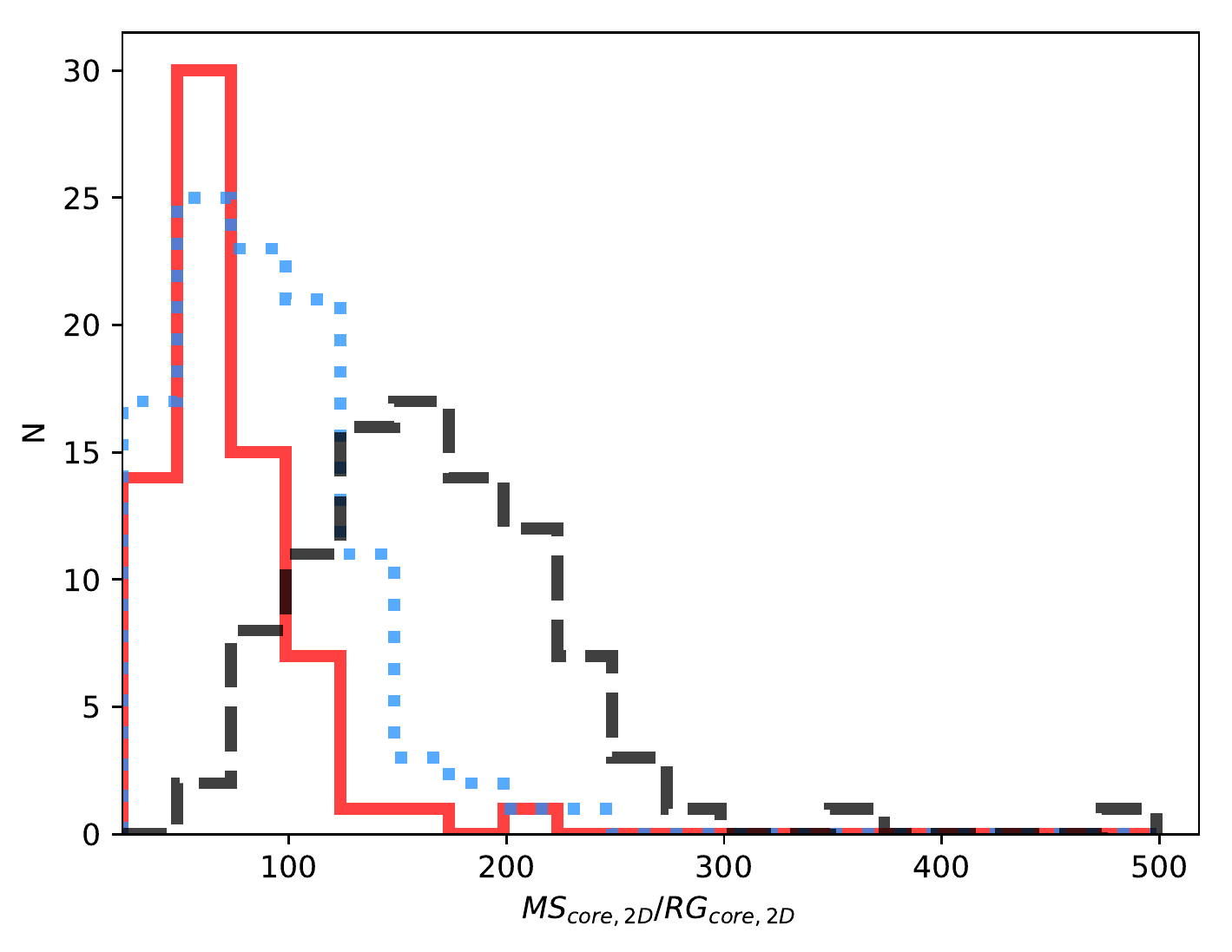}
  \end{subfigure}
  
        \caption{The ratio of the total number of MS and RG stars inside the central region and at $R_c$  for the 3D snapshot (on the left, top, and below, respectively) and for the 2D projected snapshot (on the right, top, and below, respectively) histograms. Standard models are shown in the solid red lines, IMBH are shown in the blue dashed lines,  and BHS are shown in the black dotted lines. The IMBH models show a small ratio, meanwhile, for BHS the distribution is extended to higher values.}
    \label{Fig:number-star-type}
\end{figure*}

Regarding the different central color, the BHS models show a bluer central region than the Standard and IMBH ones. This is again explained by a different dynamical history: the interactions of massive stars with IMBH would imply the removal of the former, so that mostly red stars (giant and low mass main sequence stars) would survive in the central part of the system. On the other hand, this is not strictly true for Standard and BHS models, where a higher number of main sequence stars are expected in the central part. Moreover, as discussed before, the observed core and central radius for BHS is expected to be larger than for Standard models, so the number of stars in the central part should be higher for the former. This is evident in Fig. \ref{Fig:number-star-type}, where we show the ratio of the total number of main sequence (MS) and red giant (RG) stars inside the central region and the $R_c$ , for the 3D snapshots and the projected 2D snapshots. The mean number of MS and RG stars inside the considered regions, for different dynamical models, are shown in Table \ref{Table:mean-type}. It is clear that the central region of IMBH models contains a smaller mean number of MS stars compared to the mean number of RG stars, but the Standard and BHS models have on average an higher number of MS compared to the RG stars. 

\begin{table}
    \centering
    \begin{tabular}{ cccccc}
    \hline
    & Region & Star Type & Standard & IMBH & BHS\\
    \hline
    \multirow{4}{*}{3D} &\multirow{2}{*}{Central} & MS & 2200 & 94 & 2208\\
      &  & RG & 60 & 3 & 29  \\
         \cline{2-6}
     & \multirow{2}{*}{Core} & MS & 39715 & 870 & 92062 \\
      &  & RG & 620 & 19 & 666 \\
      \hline
    \multirow{4}{*}{2D} &\multirow{2}{*}{Central} & MS & 1624 & 1201 & 4325\\
      &  & RG & 29 & 23  & 31 \\
         \cline{2-6}
     & \multirow{2}{*}{Core} & MS &26870 & 25963 & 111870\\
      &  & RG & 379 & 295 & 637 \\
      \hline
    \end{tabular}
    \caption{ Mean value for the number of MS and RG stars, in the 3D snapshot and in the 2D projected snapshot, for different dynamical models. The second column names the region considered, the third column the type of star. The forth, the fifth and the sixth names the mean value for Standard, IMBH, and BHS models, respectively. The mean value for each region and for each star type is calculated as the sum over all specific dynamical models, divided by the total number of models (Standard - 69; IMBH - 104; BHS - 93).}
    \label{Table:mean-type}
\end{table}

As shown in Fig. \ref{Fig:color-gradient}, the CRVD is already a powerful tool to distinguish IMBH models. However, the RVD gives the best separation only among the IMBH models and the two other dynamical ones. It actually enhances the difference in kinematics between the central part and the $R_{hl}$, with the former more important in IMBH models, due to the deep central potential. The similar value and spread of RVD for BHS and Standard, but a different one for IMBH, could mean that they belong to two different dynamical families: the presence of an IMBH and the deep central potential would lead to a completely different history and structure of the system. Meanwhile, BHS and Standard models have a similar evolutionary history, driven mostly by binary energy generation leading to a bigger and smaller core size (because BHs generate more energy than stars). Meanwhile for IMBH the evolutionary history has been driven by dynamical interactions with the IMBH. 

The best distinctions among the dynamical models are achieved when considering the central region properties (those being $R_c$, CRVD, CSB, and central colors). Even if this is not prohibitive to observe for MWGCs, it could be challenging for EGGCs. Indeed, the mean value for $R_c$  in our selected models is $1.72\,pc$ (ranging from $0.16$ up to $10.5 \,pc$), for $R_{hl}$ is $3.8\,pc$ (ranging from $0.9$ up to $10.7 \,pc$) and for the $10\%$ light radius $0.77\,pc$ (ranging from $0.002$ up to $3.78 \,pc$).

If we consider EGGCs in the Local Group\footnote{The distance for the LMC is assumed to be $50 \,kpc$, for the Andromeda galaxy $770\,kpc$, and for the Virgo Cluster $16.5 Mpc$.}, the aperture size of the Large Magellanic Cloud (LMC) for the mean value of $R_c$  is $7.09''$ (from $0.66''$ up to $43.31''$), of $R_{hl}$ is $15.67''$ (from $3.71''$ up to $44.14''$) and of the $10\%$ light radius is $3.18''$ (from $0.008''$ up to $15.59''$). The aperture size of the Andromeda galaxy distance, for the mean value of $R_c$  is $0.46''$ (from $0.04''$ up to $2.81''$), of  $R_{hl}$ is $1.02''$ (from $0.24''$ up to $2.87''$) and of the $10\%$ light radius is $0.21''$ (from $0.0005''$ up to $1.01''$). Instead, if we consider EGGCS in the Virgo cluster, the aperture size for the mean value of $R_c$  is $0.02''$ (from $0.002''$ up to $0.13''$), of $R_{hl}$ is $0.04''$ (from $0.01''$ up to $0.13''$) and of the $10\%$ light radius is $0.01''$ (from $2.5\cdot 10^{-5}\,''$ up to $0.05''$). The Hubble space telescope has a spatial resolution of $0.04 \sim 0.05''$ (so we could definitely observe the LMC, but some difficulties would arise for the $10\%$ light radius for some GCs); the VLT telescope (Narrow Field Mode of MUSE, for example, has spatial resolution of $0.055 ''$ - $0.08''$), instead, has a spatial resolution of $0.05''$, reaching a value of $ 0.002''$ when all the telescopes are combined (possible up to the Virgo distance, but maybe not for the $10\%$ light radius where we are at the extreme).

In this preliminary work, we showed that our approach is working well for MWGCs. The results shown are in good agreement with observations, and with the results of previous works. This gives confidence that it should also work for EGGCs, assuming that MWGCs' and EGGCs' formation processes are similar. Distinguishing between different evolutionary dynamical tracks for EGGCs could be also useful to better estimate the BH-BH merger rate, the number of exotic binaries in the local Universe (cataclysmic variable stars, X-ray binaries, etc), the number of tidal disruption events (TDE) around a IMBH, or the expected number of IMBH in a nuclear star cluster (NCS). In order to distinguish the type of cluster evolution, in the near future photometric (in different bands) and spectroscopic observations of the central properties of GCs would be needed. As shown, this may be possible for GCs in the Local Group, where the resolution needed to resolve the central part of the system could also be approachable given current technologies. If the central properties would actually be inaccessible with the current (or future) technologies, determining the ``average'' properties at $R_{hl}$ would be more realizable, even with current telescopes. Indeed, the ``average'' properties give an overall acceptable result, roughly dividing the different dynamical models.

\section{Summary and future work}  \label{sec:Summary}
In this work, we used models from the MOCCA-Survey Database \citep{Askar2017}. We limit ourselves to a subsample of models that would mimic the observational limitations and realistic properties for EGGCs, which are models having current luminosity $> 2  \cdot 10^4 L_\odot$, with a mass fallback prescription and which were initially tidally underfilling.
. For each model, we projected the  $12 \,\,Gyr$ snapshot, in order to determine the observed structural parameter (such as $R_c$, CSB, CRVD). The models have been divided accordingly by their dynamical state at $12\,\, Gyr$, that is if an IMBH, a BHS, or neither are present in the system.

Our main results can be summarized as follows:
\begin{itemize}
    \item in Sec. \ref{sub-sec:MWGCs} and Sec. \ref{sub-sec:comp-prev-works}, we showed that, overall, our subsample reproduces the observed properties of MWGCs, while being in rough agreement with the results of previous considered works \citep{Askar2018,Weatherford2019,ArcaSedda2019};
    \item the significance of dynamical history in color distribution could be important for the spread in metallicity but, due to the small number of our statistics, further study is needed (Sec. \ref{subsec:color});
    \item in Sec. \ref{subsec:3D}, we established that in order to differentiate between the three dynamical models, at least three observational parameters are needed. The best choice would be the central properties; however, a good result is obtained also from the properties at $R_{hl}$, which may be easier to observe in EGGCs ($R_{hl}$, Mean SB, and RVD).
\end{itemize}

Photometric and spectroscopic studies, in the central part or at $R_{hl}$ of EGGCs, are necessary to distinguish the type of cluster evolution. Current technologies could be limiting the distance of possible observations, however it could be not so prohibitive in the Local Group. Even if the number of systems is not so high, combining those observation with those from MWGC, it would be possible to check and verify the correlations between the global properties of the GCs and their internal dynamical state.

The next step in our work plan is to populate GCs around an external galaxy. The distribution of position in the galactic surroundings, age, metallicity (and other GC's properties) will be selected according to the observed distribution. Those properties depend on the host galaxy's type, mass, size, and luminosity, as well as the number of GCs surrounding the host galaxy. We will adopt the procedure used in this paper to simulate the EGGC population, using models  from the MOCCA-Survey Database. 
 For each GC's properties expected from the distribution, we will consider the model in our database that will best match them. In this way, we will obtain a simulated external galaxy and its GC population, as in real galaxies. Finally, we will recreate a mock observation and we will apply our methodology to this simulated EGGC population, in order to mimic real observations as much as possible.


\section*{Acknowledgements}
MG and AL were partially supported by the Polish National Science Center (NCN) through the grant UMO-2016/23/B/ST9/02732. The authors would like to express their gratitude to prof. E. Peng: his support and encouragement were invaluable. This manuscript was edited for English language by D. Abarca. AL would like to offer his special thanks and deepest gratitude to V. S. Iorio, a friend and a mentor. Without her encouragement and counseling, all this would not have materialized ever.

\bibliographystyle{mnras}
\bibliography{biblio}

\bsp
\label{lastpage}

\end{document}